\documentclass[nohyper]{JHEP3}

\usepackage{amsmath,amssymb}
\usepackage[dvips]{graphicx}
\usepackage{afterpage}
\usepackage{pifont}
\usepackage{array}

\newcommand{\auxNH}[3]{%
  \raisebox{-3.7mm}[0.0mm][5.0mm]{\makebox[0pt][l]{\ding{#1}}}%
  \raisebox{-0.8mm}[0.0mm][0.0mm]{\makebox[0pt][l]{\ding{#2}}}%
  \raisebox{+3.7mm}[7.5mm][0mm]{\makebox{\ding{#3}}}}

\newcommand{\auxIH}[3]{%
  \raisebox{-3.7mm}[0.0mm][5.0mm]{\makebox[0pt][l]{\ding{#1}}}%
  \raisebox{+0.8mm}[0.0mm][0.0mm]{\makebox[0pt][l]{\ding{#2}}}%
  \raisebox{+3.7mm}[7.5mm][0.0mm]{\makebox{\ding{#3}}}}

\newcommand{\auxSL}[1]{\makebox[0pt][l]{\hspace{#1}\large/}}

\newcommand{\dgN}[1]{%
  \ifnum#1=1\auxNH{182}{183}{184}\else%
  \ifnum#1=2\auxNH{182}{183}{174}\else%
  \ifnum#1=3\auxNH{182}{173}{184}\else%
  \ifnum#1=4\auxNH{172}{183}{184}\else%
  \ifnum#1=5\auxNH{172}{173}{184}\else%
  \ifnum#1=6\auxNH{172}{183}{174}\else%
  \ifnum#1=7\auxNH{182}{173}{174}\else%
  \ifnum#1=8\auxNH{172}{173}{174}\fi\fi\fi\fi\fi\fi\fi\fi}

\newcommand{\dgI}[1]{%
  \ifnum#1=1\auxIH{184}{182}{183}\else%
  \ifnum#1=2\auxIH{184}{182}{173}\else%
  \ifnum#1=3\auxIH{184}{172}{183}\else%
  \ifnum#1=4\auxIH{174}{182}{183}\else%
  \ifnum#1=5\auxIH{174}{172}{183}\else%
  \ifnum#1=6\auxIH{174}{182}{173}\else%
  \ifnum#1=7\auxIH{184}{172}{173}\else%
  \ifnum#1=8\auxIH{174}{172}{173}\fi\fi\fi\fi\fi\fi\fi\fi}

\newcommand{\mdN}[1]{%
  \ifnum#1=1{LMH}\else%
  \ifnum#1=2{LM\auxSL{1pt}H}\else%
  \ifnum#1=3{L\auxSL{1pt}MH}\else%
  \ifnum#1=4{\auxSL{0pt}LMH}\else%
  \ifnum#1=5{\auxSL{0pt}L\auxSL{1pt}MH}\else%
  \ifnum#1=6{\auxSL{0pt}LM\auxSL{1pt}H}\else%
  \ifnum#1=7{L\auxSL{1pt}M\auxSL{1pt}H}\else%
  \ifnum#1=8{\auxSL{0pt}L\auxSL{1pt}M\auxSL{1pt}H}\fi\fi\fi\fi\fi\fi\fi\fi}

\newcommand{\mdS}[1]{%
  \ifnum#1=1{123}\else%
  \ifnum#1=2{12\auxSL{-1pt}3}\else%
  \ifnum#1=3{1\auxSL{-0.5pt}23}\else%
  \ifnum#1=4{\auxSL{-0.7pt}123}\else%
  \ifnum#1=5{\auxSL{-0.7pt}1\auxSL{-0.5pt}23}\else%
  \ifnum#1=6{\auxSL{-0.7pt}12\auxSL{-1pt}3}\else%
  \ifnum#1=7{1\auxSL{-0.5pt}2\auxSL{-1pt}3}\else%
  \ifnum#1=8{\auxSL{-0.7pt}1\auxSL{-0.5pt}2\auxSL{-1pt}3}\fi\fi\fi\fi\fi\fi\fi\fi}

\title{Effect of Leptonic CP Phase in $\nu_{\mu} \to \nu_{\mu}$ Oscillations}

\author{Keiichi Kimura\\
  Department of Physics, Nagoya University, Nagoya, 464-8602, 
  Japan \\
  E-mail: \email{kimukei@eken.phys.nagoya-u.ac.jp}}

\author{Akira Takamura\\
  Department of Mathematics, Toyota National College of Technology, 
  Eisei-cho 2-1, Toyota-shi, 471-8525, Japan\\
  E-mail: \email{takamura@eken.phys.nagoya-u.ac.jp}}

\author{Tadashi Yoshikawa\\
  Department of Physics, Nagoya University, Nagoya, 464-8602, 
  Japan \\
  E-mail: \email{tadashi@eken.phys.nagoya-u.ac.jp}}

\abstract{%
In the case of large 1-3 mixing angle as $\sin^2 2\theta_{13}\geq 0.03$, 
we investigate the possibility for measuring the leptonic CP phase 
by using only $\nu_{\mu} \to \nu_{\mu}$ oscillations 
independently of $\nu_{\mu} \to \nu_e$ oscillations.
As the result, we find that 
the best setup to measure the CP phase without strongly 
depending on the uncertainties of the
other parameters is 
around the energy 
$E=0.43$GeV and the baseline length $L=5000$km. 
In this region, the CP phase effect remains even after averaging 
over the neutrino energy.
We also find that there is a CP sensitivity even in the 
short baseline length $L\leq 1000$km if $\Delta m_{31}^2$ is 
determined with an uncertainty of about $0.1\%$.
In the T2KK experiment, we explore the possibility for 
measuring the CP phase by using a baseline from Tokai to Kamioka 
after determining $\Delta m_{31}^2$ at the baseline to Korea. 
As the result, we find that some information of the CP phase 
can be obtained from the measurements.}


\begin{document}

\section{Introduction}
The finite mass of neutrinos and the mixings among different flavors 
have been confirmed in various neutrino experiments.
Recently, the LSND anomaly \cite{LSND} is refused by the result of 
MiniBOONE experiment \cite{MiniBOONE} and the results obtained in the past 
neutrino experiments can be almost explained by the neutrino oscillations among 
three generations.
The effect of such neutrino oscillations has been considered not only 
in the experiments on the earth but also in the far universe for example 
supernova explosions \cite{Sato, Ando, Yoshida}.

For the values of mass squared differences and mixing angles, 
the results of atmospheric neutrino experiments \cite{atmos}, 
K2K experiment \cite{K2K} and MINOS experiment \cite{MINOS} provide 
\begin{eqnarray}
|\Delta m^2_{31}|\sim 2.5\times 10^{-3} {\rm eV}^2, \quad
\sin^2 2\theta_{23}\sim 1
\end{eqnarray}
and the results of solar neutrino experiments \cite{solar} and 
KamLAND experiment \cite{KamLAND} provide
\begin{eqnarray}
\Delta m^2_{21}\sim 8.1\times 10^{-5} {\rm eV}^2, \quad
\sin^2 \theta_{12}\sim 0.31.
\end{eqnarray}
On the other hand, only the upper bound for the 1-3 mixing angle 
\begin{eqnarray}
\sin^2 2\theta_{13}\leq 0.16
\end{eqnarray}
is obtained from the CHOOZ experiment \cite{CHOOZ}.
We cannot determine the sign of $\Delta m^2_{31}$ at present 
from the experimental data.
Furthermore, we have no information on the leptonic CP phase $\delta$. 
Unveiling these unknown parameters is one of the most important aims 
in the next generation neutrino experiments. 
In particular, the value of $\delta$ is very important 
at the view point of the leptogenesis \cite{leptogenesis}.

One of the serious obstacles in determining the value of $\delta$ 
is the eight-fold degeneracies \cite{Burguet0103, Minakata, Barger}.
There are some cases in which the uncertainty of $\delta$ becomes large 
due to the effect of degeneracy.
One of the turning points is whether $\theta_{13}$ can be determined by 
the next generation reactor experiments like Double CHOOZ experiment 
\cite{Double CHOOZ} and the superbeam experiments like T2K experiment 
\cite{Itow}, NO$\nu$A experiment \cite{NOvA}.
In this paper, we concentrate on the cases that $\sin^2 2\theta_{13}$ is larger 
than $0.03$ and assume that $\theta_{13}$ will be found in 
the Double CHOOZ experiment without 
being affected by the $\theta$-$\delta$ ambiguity.
In this case, it is suggested by many authors that the remaining 
four-fold degeneracies can be also solved within a few decades.
One possibility is the observation of neutrinos from the same beam 
to two detectors on different baselines 
like Tokai-to-Kamioka-Korea (T2KK) proposal \cite{T2KK} and 
SuperNO$\nu$A proposal \cite{SuperNOvA}.
The other possibilities are the combination of more than two kinds of neutrino sources, 
long baseline plus atmospheric neutrinos \cite{Gandhi, LBL-atmos} and 
long baseline plus reactor neutrinos \cite{LBL-reactor}.
The possibility for using the Wide Band Beam and 
analyzing the spectral information of neutrino events is also investigated 
in ref. \cite{WBB}.
In these proposals, the measurement of the leptonic CP phase is also explored 
by using $\nu_{\mu} \to \nu_e$ ($\nu_e \to \nu_{\mu}$) oscillations.

In our previous work \cite{Kimura0603}, 
we suggested that we can measure the leptonic CP phase by using only 
$\nu_{\mu} \to \nu_{\mu}$ oscillations in the region $E\leq 2$GeV, $L\geq 2000$km 
if $\sin^2 2\theta_{13}$ is large.
In the analysis, the probability for $\nu_{\mu} \to \nu_{\mu}$ oscillations 
can change about $0.4$ by the CP phase effect and then there remains a difference of 
about $0.2$ between the maximal and minimal values of the probabilities 
even after averaging over the neutrino energy.
In these considerations, we have explored new possibilities of experiments to be 
performed after a decade in addition to solving parameter degeneracies and 
determining the value of the CP phase.  
It is easy to observe the $\nu_\mu \to \nu_\mu $ oscillations in superbeam 
experiments and neutrino factory experiments.
There is a potential to measure the CP phase $\delta$ independently of  
$\nu_{\mu} \to \nu_e$ oscillations in the case that the parameters except for $\delta$ 
can be measured precisely.
If we assume the unitarity in three generations, the channel of 
$\nu_{\mu} \to \nu_e$ oscillations should be related to that of 
$\nu_{\mu}\to \nu_{\mu}$ oscillations.
This means that we can predict the behavior of one oscillation channel by 
the behavior of the other channel.
(See references \cite{unitarity} about the discussion of the unitarity 
in the lepton sector, for example.) 
If we find the difference between this prediction and the experimental 
result, we must consider effect of new physics and we will have 
some constraints to the unified theory in the high energy physics. 
Thus, the measurements of $\delta$ in two independent channels 
are very important for exploring the new physics beyond the 
Standard model.
Recently, the exact formulation of neutrino oscillation probabilities 
has been extended to the case of non-standard interaction 
in view of the era of precision measurement of parameters \cite{Yasuda}.
A lot of investigations have been also performed 
about the exploration of non-standard interaction 
and the non-unitary effect by future experiments.
See \cite{non-standard}, \cite{non-unitary} and the references therein.

In this paper, we give the detailed analysis for the possibility of 
the measurement of the CP phase in $\nu_{\mu}\to \nu_{\mu}$ oscillations, 
considering the systematic error, 
uncertainties of parameters except for $\delta$ and 
the uncertainty of matter density, which were not considered 
in our previous work \cite{Kimura0603}.
We investigate the best conditions of baseline length and energy 
for measuring the CP phase in $\nu_{\mu}\to \nu_{\mu}$ oscillations 
by using both analytical and numerical methods.
We also calculate the baseline dependence and $\theta_{13}$ dependence 
of CP sensitivity.
Furthermore, we consider the reason why we have no CP sensitivity 
in comparatively short baseline length and show how the 
CP sensitivity depends on the uncertainty of $\Delta m_{31}^2$. 
We estimate how this uncertainty can be improved in future experiments 
and how CP sensitivity can become good 
in comparatively short baseline experiments.
We usually use the $\nu_{\mu}\to \nu_{\mu}$ channel in order to 
measure $\Delta m_{31}^2$ precisely.
Namely, we need two different baselines to measure the 
CP phase in the case of short baseline by using this channel. 
So, we consider the T2KK experiment as concrete setup.
We use one baseline from Tokai to Korea to measure the value of 
$\Delta m_{31}^2$ as precise as possible and use another baseline 
from Tokai to Kamioka to measure the CP phase.

Outline of this paper is the following.
In section 2, we explore the energy and the baseline length 
where the CP dependence in $\nu_{\mu}\to \nu_{\mu}$ oscillations 
becomes large.
In section 3, we assume the concrete experimental setup and 
calculate the CP sensitivity.
In section 4, we clarify what is the problem for measuring the 
CP phase in relatively short baseline.
Then, in section 5, we reinvestigate the CP sensitivity 
in the case that this problem can be improved.
In section 6, we conclude.
Finally, in appendix, the approximate formula for the 
coefficient of $\cos \delta$ is derived, 
which is applicable for the case of large $L/E$.

\section{Region with Large CP Dependence}
In this section, let us review the CP dependence of 
neutrino oscillation probabilities observed in the superbeam 
experiments.
We also investigate the energy and baseline length where the 
CP dependence becomes large.

The Hamiltonian in matter is represented as 
\begin{equation}
H = O_{23} \Gamma H'\Gamma^\dagger O_{23}^T,
\end{equation}
where $O_{23}$ is the rotation matrix between the second and the third 
generations and $\Gamma$ is the phase matrix defined by 
$\Gamma={\rm diag}(1,1,e^{i\delta})$.
Without loss of generality, we can factor out the part of $\theta_{23}$ and 
$\delta$.
This makes it possible to include the matter effect only in the reduced 
Hamiltonian $H^{\prime}$ as 
\begin{eqnarray}
H^{\prime}=O_{13}O_{12}{\rm diag}(0,\Delta_{21},\Delta_{31})O_{12}^TO_{13}^T
+ {\rm diag}(a,0,0).
\end{eqnarray}
Due to this factorization, we can transparently understand 
how oscillation probabilities depend on the CP phase $\delta$.
In the above expression, we use the equalities $\Delta_{ij}=\Delta m_{ij}^2/(2E)=(m_i^2-m_j^2)/(2E)$,
$a=\sqrt2 G_F N_e\simeq 7.56\times 10^{-5}\cdot \rho Y_e$, 
$G_F$ is the Fermi constant, $N_e$ is the electron number density, $\rho$ is the 
matter density, $Y_e$ is the fraction of electrons, $E$ is the 
neutrino energy and $m_i$ is the mass of $\nu_i$.
The CP dependences of the probabilities for $\nu_{\mu} \to \nu_e$ oscillation, 
$\nu_{\mu} \to \nu_{\mu}$ oscillation and $\nu_{\mu}\to \nu_{\tau}$ oscillation 
are given by 
\begin{eqnarray}
P_{\mu e}&=&A_{\mu e}\cos \delta+B_{\mu e}\sin \delta+C_{\mu e}, \\
P_{\mu\mu}&\simeq&A_{\mu\mu}\cos \delta\hspace{1.7cm}+C_{\mu\mu}, \label{7}\\
P_{\mu\tau}&\simeq&\hspace{1.7cm}B_{\mu \tau}\sin \delta+C_{\mu\tau},
\label{8}
\end{eqnarray}
in ref. \cite{Yokomakura}. Eqs. (\ref{7}) and (\ref{8}) hold exactly in the case of 
$\theta_{23}=45^{\circ}$ \cite{Takamura0403}, where the coefficients 
$A_{\mu\alpha}$, $B_{\mu\alpha}$ and $C_{\mu\alpha}$ $(\alpha=e,\mu,\tau)$ are  
the quantities determined by the parameters except for $\delta$.
If we assume the unitarity in the framework of three generations, 
the sum of these probabilities has to be one for any value of $\delta$.
This leads to the relations among the coefficients 
\begin{eqnarray}
&&A_{\mu e}+A_{\mu\mu}\simeq 0, \\
&&B_{\mu e}+B_{\mu\tau}= 0, \\
&&C_{\mu e}+C_{\mu\mu}+C_{\mu\tau}=1.
\end{eqnarray}
It is well known that the order of magnitude for the coefficients is represented as 
\begin{eqnarray}
&&A_{\mu\mu}=-A_{\mu e}=O(\alpha s_{13}), \qquad 
B_{\mu\tau}=-A_{\mu e}=O(\alpha s_{13}), \\
&&C_{\mu e}=O(\alpha^2)+O(s_{13}^2), \qquad 
C_{\mu\mu}=O(1), \qquad C_{\mu\tau}=O(1), 
\end{eqnarray}
by using the small parameters $\alpha=\Delta_{21}/\Delta_{31}$ and 
$\sin \theta_{13}=s_{13}$.
See ref. \cite{Takamura0403} for example.
It should be noted that the magnitude of $A_{\mu\mu}$ is 
as same as that of $A_{\mu e}$ and is not so small.
Therefore, $\nu_{\mu} \to \nu_{\mu}$ oscillation 
will be one of the important channels in order to obtain the information on the 
CP phase, although it is hard to observe the CP phase 
compared to $\nu_{\mu} \to \nu_e$ channel because of the large CP independent 
term $C_{\mu\mu}$.
If we observe some differences between the values of the CP phase 
measured by the two independent channels, this means the violation 
of the unitarity in three generations and we can obtain some of the important 
clues for new physics.

Below, it is considered how we should choose the energy and the baseline 
in order to measure the CP phase by using only $\nu_{\mu}\to \nu_{\mu}$ 
oscillations.
At first, we numerically calculate the region in $E$-$L$ plane 
with large $|A_{\mu\mu}|$, which is the coefficient of $\cos \delta$. 
In this calculation, we use the following parameters, 
$\Delta m^2_{31}=2.5\times 10^{-3} {\rm eV}^2$, 
$\sin^2 2\theta_{23}=1$, 
$\Delta m^2_{21}=8.1\times 10^{-5} {\rm eV}^2$, 
$\sin^2 \theta_{12}=0.31$, $\sin^2 2\theta_{13}=0.16$, $\rho=3.3$g/cm$^3$ 
and $Y_e=0.494$. 

In figure 1, the black color shows the region with large $|A_{\mu\mu}|$, 
namely at relatively low energy $E\leq 2$GeV and a long baseline $L\geq 2000$km.
In other words, $|A_{\mu\mu}|$ becomes large in the region 
with approximately $L({\rm km})/E({\rm GeV})\geq 2000$.
However, in present experiments, only the event rate averaging over the energy 
can be observed in the region with such large $L/E$ due to the finite energy 
resolution of the detector.
Hence, we need to learn whether the CP phase effect remains in such situation 
and we would like to investigate the condition 
under which the CP phase effect becomes the largest after averaging.

In order to investigate the behavior of $A_{\mu\mu}$ more accurately, 
we derived the approximate formula of $A_{\mu\mu}$ from the exact one 
\cite{Kimura0203} in the Appendix.
In the derivation, we kept in mind that our approximation 
should be valid for the energy region 
$E\simeq 0.1$-$10$GeV realized in superbeam experiments.
More concretely, we defined $\lambda_i$ as the effective mass of $i$-th neutrino 
divided by $(2E)$ and we took the approximation $\lambda_1<\lambda_2\ll \lambda_3$, 
$a\ll \lambda_3$ and $s_{13}^2\ll 1$. 
Then, we have left only leading order terms of small quantities, 
$\Delta_{21}$, $\lambda_1$, $\lambda_2$ and $s_{13}$.
In order to be a good approximation for the region with large $L/E$, 
we did not neglect the term with the order of $O(\Delta_{21}^{\prime})$ 
included in the oscillating part and multiplied by $L/E$.

\FIGURE[!t]{
  \includegraphics[width=0.6\textwidth]{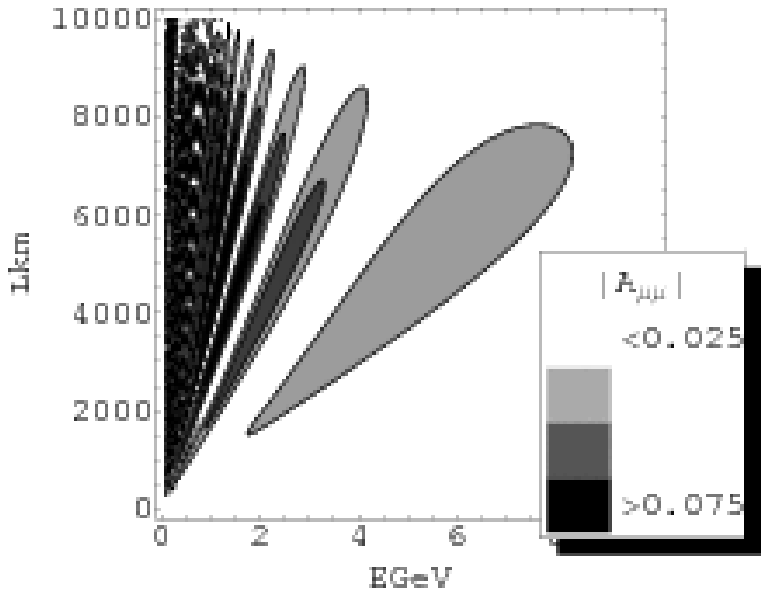}
  \caption{\label{fig:fig1}%
    Region with large $|A_{\mu\mu}|$.
In the black region, the magnitude of $|A_{\mu\mu}|$ becomes large.}}

Hence, the approximate formula for $A_{\mu\mu}$ is calculated as 
\begin{eqnarray}
A_{\mu\mu}&\simeq&
\underbrace{\frac{4J_r\Delta_{21}(a-\Delta_{21}\cos 2\theta_{12})}{\tilde{\Delta}_{21}^2}
\sin^2 \tilde{\Delta}_{21}^{\prime}}_{\displaystyle{A_1}}
\underbrace{-\frac{4J_r\Delta_{21}}{\tilde{\Delta}_{21}}\sin \tilde{\Delta}_{21}^{\prime}
\sin (2\tilde{\Delta}_{31}^{\prime}-\tilde{\Delta}_{21}^{\prime})}_{\displaystyle{A_2}}
\label{Ammapp},
\end{eqnarray}
where $J_r=s_{12}c_{12}s_{23}c_{23}s_{13}c_{13}^2$, $\tilde{\Delta}_{ij}=\lambda_i-\lambda_j$, 
$\tilde{\Delta}_{ij}^{\prime}=\tilde{\Delta}_{ij}L/2$ and 
\begin{eqnarray}
\lambda_1&\simeq&\frac{\Delta_{21}+a
-\sqrt{(a-\Delta_{21}\cos 2\theta_{12})^2+\Delta_{21}^2\sin^2 2\theta_{12}}}{2}\\
\lambda_2&\simeq&\frac{\Delta_{21}+a
+\sqrt{(a-\Delta_{21}\cos 2\theta_{12})^2+\Delta_{21}^2\sin^2 2\theta_{12}}}{2}\\
\lambda_3&\simeq&\Delta_{31}.
\end{eqnarray}
In eq.(\ref{Ammapp}), $A_{\mu\mu}$ is represented as the sum of two terms $A_1$ and $A_2$.
$A_1$ is slowly oscillating term according to the change of energy as controlled by 
$\tilde{\Delta}_{21}^{\prime}$. $A_2$ is rapidly oscillating term as controlled by 
$\tilde{\Delta}_{31}^{\prime}$.
In the small $L/E$ region, $A_1$ can be neglected and the main contribution comes from $A_2$.
As the value of $L/E$ increases, $A_1$ also gives the contribution and 
$A_2$ oscillates faster. 
Therefore, only $A_1$ remains in the region with sufficiently large $L/E$ 
and after averaging over the energy.
The total behavior of $A_{\mu\mu}$ can be described as the oscillation 
around the average value determined by $A_1$.
The coefficient of sine function in $A_1$ is given by 
\begin{eqnarray}
\frac{4J_r\Delta_{21}(a-\Delta_{21}\cos 2\theta_{12})}{\tilde{\Delta}_{21}^2}
=\frac{4J_r\Delta m^2_{21}(2aE-\Delta m^2_{21}\cos 2\theta_{12})}{(2aE-\Delta m^2_{21}\cos 2\theta_{12})^2
+\Delta m_{21}^4\sin^2 2\theta_{12}}.
\end{eqnarray}
If we use the parameters 
$\sin^2 2\theta_{23}=1$ and $\sin^2 \theta_{12}=0.31$, it is found that 
the value of local maximum is given by 
\begin{eqnarray}
A_1^{\rm max}=\frac{2J_r}{\sin 2\theta_{12}}\sin^2 
\left(\frac{\sqrt{2}\Delta m^2_{21}\sin 2\theta_{12}L}{4E_{\ell}}\right)
=\frac{\sin 2\theta_{13}}{4}\sin^2 \left(\frac{\sqrt{2}\Delta m^2_{21}\sin 2\theta_{12}L}{4E_{\ell}}\right)
\label{2.20}
\end{eqnarray}
at the energy 
\begin{eqnarray}
E_{\ell}=\frac{\Delta m^2_{21}(\cos 2\theta_{12}+\sin 2\theta_{12})}{2a}
=0.43{\rm GeV}\cdot\frac{\Delta m^2_{21}}{8.1\times 10^{-5} {\rm eV}^2}\cdot \frac{3.3{\rm g/cm}^3}{\rho}. 
\label{2.21}
\end{eqnarray}
If we fix the energy at this value and from the maximal condition 
$\sqrt{2}\Delta m^2_{21}\sin 2\theta_{12}L/4E_{\ell}=(2n+1)\pi/2$ $(n=0,1,2,\cdots)$ for (\ref{2.20}),
we can determine the baseline length $L_{\ell}$ as 
\begin{eqnarray}
L_{\ell}=\frac{\sqrt{2}E_{\ell}(2n+1)\pi}{\Delta m^2_{21}\sin 2\theta_{12}}
=5000{\rm km}\cdot (2n+1)\cdot \frac{8.1\times 10^{-5} {\rm eV}^2}{\Delta m^2_{21}}\cdot \frac{\rho}{3.3{\rm g/cm}^3}.
\label{25}
\end{eqnarray}
If we use the average density calculated in the PREM \cite{PREM} corresponding to each baseline,
$A_1$ becomes maximal at $L_{\ell}=5000$km and $10000$km in the earth mantle.
We also find from (\ref{25}) that $A_1^{\rm max}$ attains to about $0.1$ in the case of 
$\sin^2 2\theta_{13}=0.16$.

Next, let us consider $A_2$.
The factor $(4J_r\Delta_{21}/\tilde{\Delta}_{21})\sin \tilde{\Delta}_{21}^{\prime}$ included in $A_2$ 
does not change so much compared with $\sin (2\tilde{\Delta}_{31}^{\prime}-\tilde{\Delta}_{21}^{\prime})$ 
for the change of energy.
Namely, we regard 
\begin{eqnarray}
\frac{4J_r\Delta_{21}}{\tilde{\Delta}_{21}}\sin \tilde{\Delta}_{21}^{\prime}
\simeq \frac{2J_r\Delta m_{21}^2}{aE}\sin (aL)
\end{eqnarray}
as the amplitude of $A_2$ and roughly speaking, the magnitude decreases inversely proportional to 
the energy.
Next, let us consider the oscillating term related to $\Delta_{31}^{\prime}$.
The constructive interference of $A_1$ and $A_2$ occurs in the case of 
$\sin (2\tilde{\Delta}_{31}^{\prime}-\tilde{\Delta}_{21}^{\prime})
\simeq \sin (2\Delta_{31}-a)^{\prime}\simeq -1$.
This leads the maximal condition $(2\Delta m^2_{31}-2aE)L/(4E)=(4n-1)\pi/2$.
In other words, $A_{\mu\mu}$ becomes maximal near the energy 
\begin{eqnarray}
E_n({\rm GeV})=\frac{4\cdot 1.27\frac{\Delta m^2_{31}}{{\rm eV}^2}\frac{L}{\rm km}}
{(4n-1)\pi+2\cdot 2\cdot1.27a\frac{L}{\rm km}}
\quad (n=1,2,3,\cdots).
\end{eqnarray}
{}From this expression, it is expected that the peak 
appears near $E=6.7, 2.9, 1.8 \cdots$ GeV in the baseline $L=5000$km.
In the case of small energy, we can approximate as 
\begin{eqnarray}
E_n({\rm GeV})\simeq \frac{4\cdot 1.27\frac{\Delta m^2_{31}}
{{\rm eV}^2}\frac{L}{\rm km}}{(4n-1)\pi}
\quad (n=1,2,3,\cdots),
\end{eqnarray}
then the difference between the neighboring peaks is given by 
\begin{eqnarray}
\Delta E({\rm GeV})=E_{n}-E_{n+1}
\simeq \frac{1.27\frac{\Delta m^2_{31}}{{\rm eV}^2}\frac{L}{\rm km}}{n^2\pi}
\simeq \frac{3.0\cdot \frac{\Delta m^2_{31}}{2.5\cdot 10^{-3}{\rm eV}^2}\frac{L}{5000{\rm km}}}{n^2},
\end{eqnarray}
at the region satisfying the condition $n\gg 1$. 
Roughly speaking, we can observe the oscillation 
in the case that the energy resolution $\sigma_E$ is smaller than 
the difference between the neighboring peaks.
For example, the energy resolution is given by $\sigma_e=0.085$GeV 
in the case of Water Cherenkov (WC) detector used in the Super-Kamiokande (SK) experiment.
{}From the above condition $\Delta E>\sigma_e$, we obtain the condition $n<6$.
This means that we can distinguish up to the fifth peak ($E\simeq
1$GeV) at the baseline length of $L=5000$km.
In the energy lower than $E\simeq 1$GeV, we cannot distinguish the different 
peaks and have to take the average.
Namely, the effect of $A_2$ can be neglected.
In figure 2, left and right figures show the magnitude of $A_{\mu\mu}$ and $P_{\mu\mu}$ 
as the function of energy at
the baseline $L=295$km and $5000$km .

In the left figures, the blue and the red lines represent the magnitudes of 
$A_{\mu\mu}$ and $A_1$.
In the right figures, the red and the blue lines are corresponding to the 
probabilities in the case of $\delta=0^{\circ}$ and $\delta=180^{\circ}$ 
respectively.
One can see that the magnitude of $A_{\mu\mu}$ is small in the case of 
$L=295$km from the top left figure.
On the other hand, the CP phase effect is large in the case of $L=5000$km 
even after the averaging and the value becomes maximal around $E=0.43$GeV. 
This coincides with the result obtained by the analytical expression.
We also find that the peaks appear in the position calculated by $A_2$ 
as we discussed before.
In the middle right figure, we have large CP dependence also in the 
survival probability.

Next, let us consider the case of inverted hierarchy.
We obtain the coefficient of $\cos \delta$ in the case of inverted hierarchy 
by the replacement $\Delta_{31} \to -\Delta_{31}$ in (\ref{Ammapp}) 
(exactly speaking, we perform the replacement 
$\Delta_{31} \to -\Delta_{31}+2\Delta_{21}\cos 2\theta_{12}$).
$A_1$ does not change in this replacement.
Therefore, the total event rates in both hierarchies are almost the same 
in the low energy region.
On the other hand, $A_{\mu\mu}$ of inverted hierarchy becomes slightly smaller 
in high energy region because of the MSW effect \cite{MSW} due to the 1-3 mixing.
Totally, the CP sensitivity is expected to be small in the case of inverted hierarchy 
compared with the case of normal hierarchy.

Here, we comment the case of anti-neutrino oscillations.
We obtain the probability $P_{\bar{\mu}\bar{\mu}}$ for the anti-neutrino oscillations 
$\bar{\nu}_{\mu} \to \bar{\nu}_{\mu}$ by the replacement $\delta \to -\delta$ and $a\to -a$ 
in $P_{\mu\mu}$.
As there exist only CP even terms in $P_{\mu\mu}$, we only have to replace 
$a$ to $-a$.
We can obtain the energy $\bar{E}_{\ell}$, where $A_1$ is maximal, as
\begin{eqnarray}
\bar{E}_{\ell}=\frac{\Delta m^2_{21}(-\cos 2\theta_{12}+\sin 2\theta_{12})}{k}
=0.17{\rm GeV}\cdot\frac{\Delta m^2_{21}}{8.1\times 10^{-5} {\rm eV}^2}\cdot \frac{3.3{\rm g/cm}^3}{\rho}, 
\end{eqnarray}
by the same procedure performed in $P_{\mu\mu}$.
However, $\bar{E}_{\ell}$ is nearly the production energy of $\mu$ and 
so it would be hard to observe the CP phase effect for this energy.
So, we have to measure the CP phase in higher energy region than $\bar{E}_{\ell}$ 
for the case of anti-neutrinos.
Considering also the smallness of the cross section of anti-neutrino, 
which is about $1/3$ of that of neutrino, the use of neutrinos has an advantage 
compared with the use of anti-neutrinos for the measurement of the CP phase 
through $\nu_{\mu}$ ($\bar{\nu}_{\mu}$) events.
In figure 3, we show the energy dependence of $A_{\bar{\mu}\bar{\mu}}$ and 
$P_{\bar{\mu}\bar{\mu}}$ in the baseline of $L=5000$km in left and right figures. 
In the left figure, the blue and the red lines represent the value of 
$A_{\bar{\mu}\bar{\mu}}$ and $A_1$.
In the right figure, the red and the blue lines correspond to the probabilities 
in case of $\delta=0^{\circ}$ and $\delta=180^{\circ}$ as same as in figure 2.
We can see that the absolute value of $A_1$ attains up to $0.1$ around the energy 
$E=0.1$-$0.2$GeV.
We also find that the sign of $A_{\bar{\mu}\bar{\mu}}$ is negative over the entire 
region. This comes from the denominator of $A_1$, which has negative sign 
for the replacement of $a\to -a$

\FIGURE[!t]{
\begin{tabular}{cc}
$A_{\mu\mu}$ \quad $L=295$km \quad normal 
& $P_{\mu\mu}$ \quad $L=295$km \quad normal \\
\hspace{-1cm}
\resizebox{72mm}{!}{\includegraphics{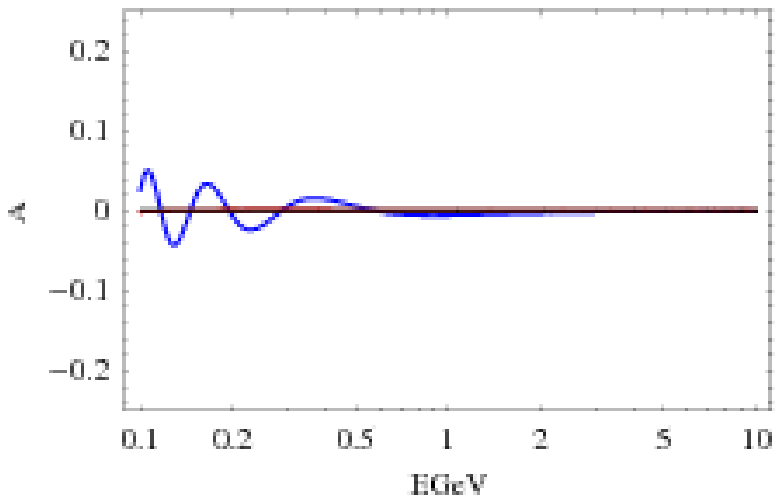}} &
\hspace{-0.5cm}
\resizebox{75mm}{!}{\includegraphics{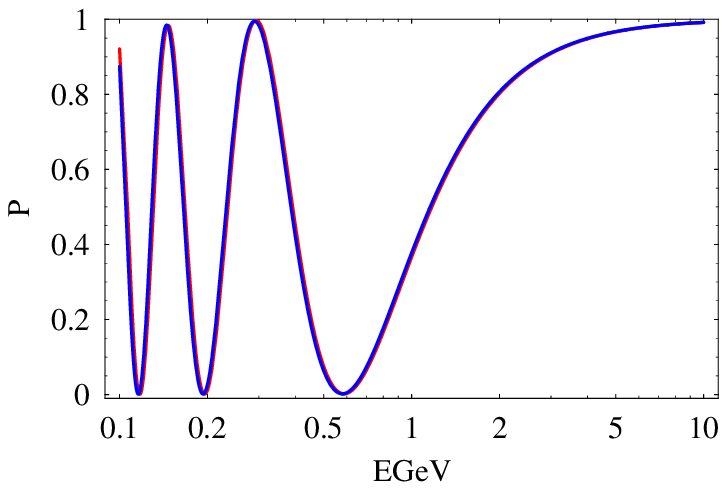}} \\
$A_{\mu\mu}$ \quad $L=5000$km \quad normal 
& $P_{\mu\mu}$ \quad $L=5000$km \quad normal \\
\hspace{-1cm}
\resizebox{72mm}{!}{\includegraphics{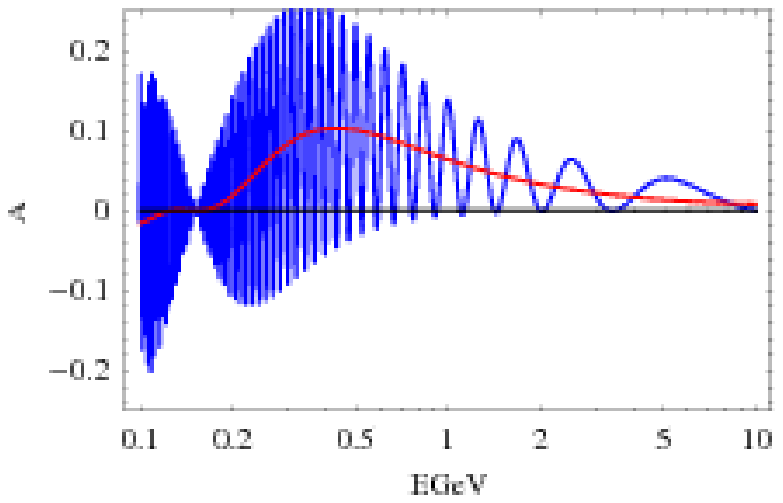}} &
\hspace{-0.5cm}
\resizebox{75mm}{!}{\includegraphics{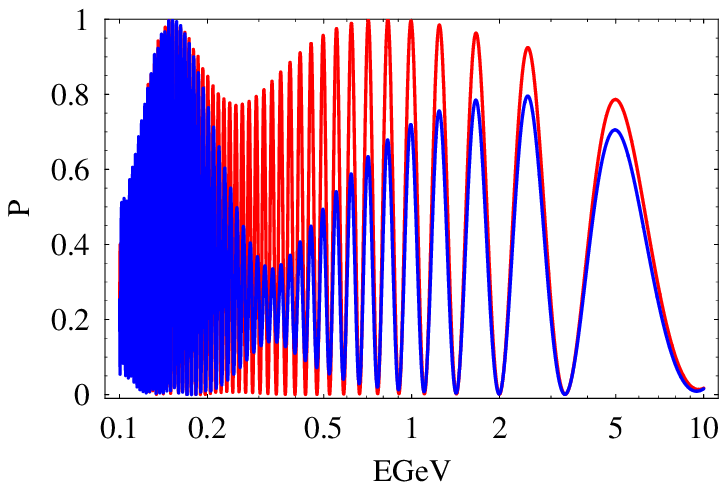}} \\
$A_{\mu\mu}$ \quad $L=5000$km \quad inverted 
& $P_{\mu\mu}$ \quad $L=5000$km \quad inverted \\
\hspace{-1cm}
\resizebox{72mm}{!}{\includegraphics{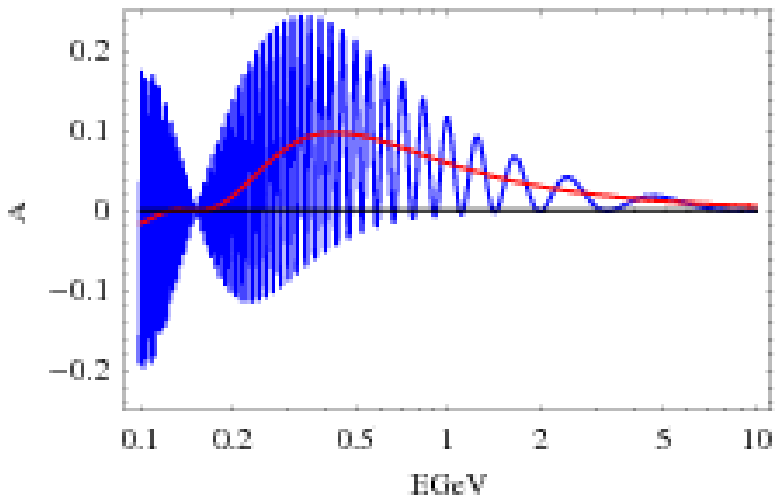}} &
\hspace{-0.5cm}
\resizebox{75mm}{!}{\includegraphics{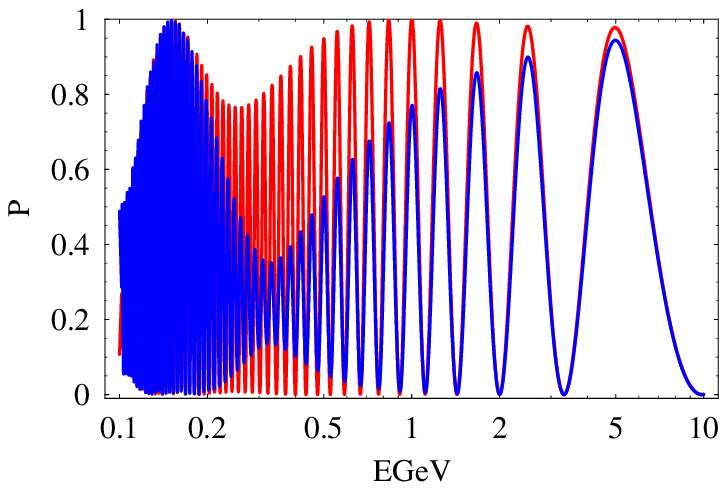}} \\

\end{tabular}
  \caption{\label{fig:fig2}%
Energy dependence of $A_{\mu\mu}$ and $P_{\mu\mu}$.
Left and right figures show the magnitude of $A_{\mu\mu}$ and $P_{\mu\mu}$.
In the left figure, blue and red lines represent the magnitudes of $A_{\mu\mu}$ 
and $A_1$ respectively. In the right figure, 
red and blue lines correspond to the true values of $\delta=0^{\circ}$ and $180^{\circ}$.
}}

\FIGURE[!t]{
\begin{tabular}{cc}
$A_{\bar{\mu}\bar{\mu}}$ \quad anti-neutrino & $P_{\bar{\mu}\bar{\mu}}$ \quad anti-neutrino \\
\hspace{-1cm}
\resizebox{72mm}{!}{\includegraphics{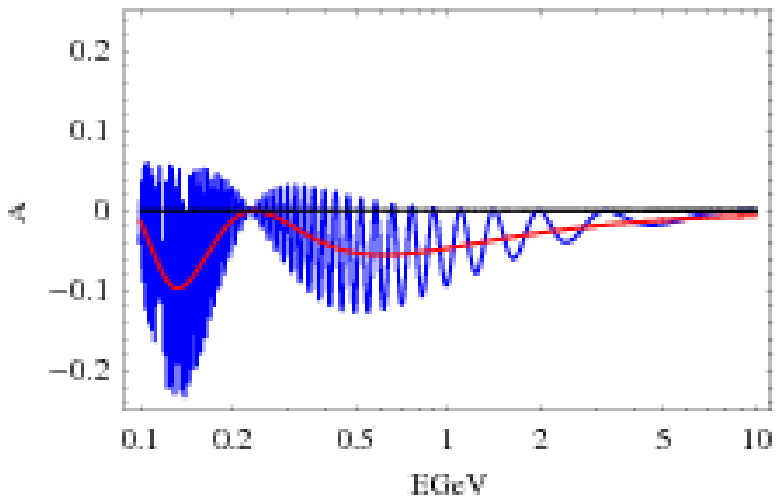}} &
\hspace{-0.5cm}
\resizebox{75mm}{!}{\includegraphics{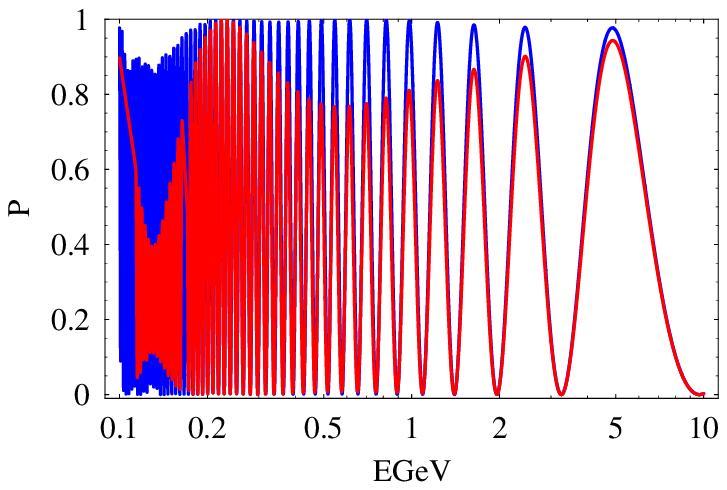}} 
\end{tabular}
  \caption{\label{fig:fig3}%
Energy dependence of $A_{\bar{\mu}\bar{\mu}}$ and $P_{\bar{\mu}\bar{\mu}}$.
Left and right figures show the magnitude of $A_{\bar{\mu}\bar{\mu}}$ 
and $P_{\bar{\mu}\bar{\mu}}$.
Blue and red lines represent the values of $A_{\bar{\mu}\bar{\mu}}$ and $A_1$
respectively in left figure. 
On the other hand, red and blue lines in right figure represent the probabilities with 
$\delta=0^{\circ}$ and $180^{\circ}$
}}

Let us summarize the results obtained in this section.
\begin{itemize}
\item In $\nu_{\mu}\to \nu_{\mu}$ oscillations, the averaged value of $A_{\mu\mu}$ 
becomes maximal around the energy $E=0.43$GeV and the baseline length 
$L=5000$km and $10000$km in the earth mantle.
\item The average value of $A_{\mu\mu}$ is positive in the region $E\geq 0.2$GeV. 
This means that we only have to measure the total rate of $\nu_{\mu}$ events 
in order to observe the CP phase effect and we need not the good energy resolution 
of the detector.
\item We obtain more information 
on the CP phase by observing the energy dependence of $\nu_{\mu}$ events at $E\geq 1$GeV
in addition to the total rate.
\item In relatively short baseline like $L=295$km, the magnitude of $A_{\mu\mu}$ becomes 
small and the observation of the CP phase effect is difficult unless 
other parameters except for $\delta$ are precisely known.
\item In the case of inverted hierarchy, the CP phase effect included in the total rate 
is similar to the case of normal hierarchy. 
The CP dependence in the high energy region is slightly reduced.
\item Considering the production energy of $\mu$ and the cross
section,
it is easy to observe the CP phase in the experiment 
by neutrinos compared with anti-neutrinos.
\item As the CP phase effect is proportional to $s_{13}$, it is difficult to observe 
in the case of small $s_{13}$.
\end{itemize}

In the above discussion, we do not consider the decrease of the flux of 
neutrinos according to the distance.
{}From the statistical point of view, the short baseline is advantageous 
because of the large event numbers.
However, in the case that the ratio of $A_{\mu\mu}$ to the probability 
is small, the observation of $\delta$ is strongly affected by the uncertainties 
of other parameters except for $\delta$ as we will discuss later.

\section{Estimation of Signal from Leptonic CP Phase}
In this section, we estimate how precise the CP phase can be measured 
in $\nu_{\mu}\to \nu_{\mu}$ oscillations by using $\chi^2$ method.
As an experimental setup, we consider the 4MW and $2.5^{\circ}$ Off-Axis 
JPARC beam and WC detector with the fiducial mass of 500kt \cite{Itow}, 
which is the same as those of T2HK experiment.
We take two kinds of baseline length $L=295$km and $5000$km 
as in the previous section.
We assume ten years data acquisition by using only neutrinos.
We use the same parameters as in figure 1.
We also assume that the mass hierarchy has already been determined before 
measuring of the CP phase in $\nu_{\mu}\to \nu_{\mu}$ oscillations.
For example, the possibility of determining the mass hierarchy by the atmospheric neutrinos
was discussed in ref. \cite{Gandhi}. 
They concluded that the observations using the $545$kt WC detector in three years 
determines the mass hierarchy at 2-$\sigma$ C.L. if $\sin^2 2\theta_{13}\geq 0.05$.
The parameter uncertainties, except for $\delta$, are assumed as 5\% 
for $\theta_{12}$, $\theta_{23}$ and $\theta_{13}$, and as 
4\% for $\Delta m^2_{21}$, and 1\% for $\Delta m^2_{31}$ by 
expecting an improvement for the next ten years \cite{Schwetz0510}. 
We also consider 5\% uncertainty of matter density.
The energy window for our analysis is $E=0.4$-$1.2$GeV and is divided into 20 bins.
We calculate $\Delta \chi^2$ by using the energy dependence of QE $\nu_{\mu}$ events 
and the total rate of CC $\nu_{\mu}$ events.
As the backgrounds, we consider NC $\nu_x$ events, where $x$ takes all flavors.
We set the uncertainties of signal and the
background normalization $n_s$, $n_b$ as 
$\sigma(n_s^{CC})=2.5\%$, $\sigma(n_s^{QE})=\infty$, $\sigma(n_b)=20\%$ 
and the uncertainties of their tilts $t_s$, $t_b$ as $\sigma(t_s)=2\%$, $\sigma(t_b)=2\%$.
As the value of $\sigma(n_s^{CC})=2.5\%$ may be optimistic, 
we discuss how the change of signal normalization affects the results later.
The reason for setting $\sigma(n_s^{QE})=\infty$ is to be normalization free for 
QE events and to prevent the double counting in QE and CC events.
We use common normalizations and tilts except for this.
The expected number of QE events $N_{QE i}$ observed in the i-th bin and CC events $N_{CC}$ 
are calculated as 
\begin{equation}
N_{QE i}=s^{QE}_i\left(1+n^{QE}_s+t^{QE}_s\cdot \frac{E_i-\bar{E}}{E_{max}-E_{min}}\right)
+b^{QE}_i\left(1+n^{QE}_b+t^{QE}_b\cdot \frac{E_i-\bar{E}}{E_{max}-E_{min}}\right), 
\end{equation}

\vspace{-0.5cm}
\begin{equation}
N_{CC}=s^{CC}\left(1+n^{CC}_s+t^{CC}_s\cdot \frac{E_i-\bar{E}}{E_{max}-E_{min}}\right)
+b^{CC}\left(1+n^{CC}_b+t^{CC}_b\cdot \frac{E_i-\bar{E}}{E_{max}-E_{min}}\right),
\end{equation}
where $s$ and $b$ are the signal and background for the case that the uncertainties 
of normalization and tilt are not considered.
See \cite{Huber0204} as the more detail of the definitions.
Furthermore, we define $\Delta \chi^2$ as 
\begin{eqnarray}
\Delta
\chi^2&=&\sum_{i=1}^{20}\frac{(N_{QE i}-N_{QE i}^{true})^2}{N_{QE i}^{true}}
+\frac{(N_{CC}-N_{CC}^{true})^2}{N_{CC}^{true}}
+\Delta \chi_{sys}^2+\Delta \chi_{para}^2, \\
\Delta \chi_{sys}^2&=&\sum_{\alpha}\left[\left(\frac{n_s^{\alpha}}{\sigma(n_s^{\alpha})}\right)^2
+\left(\frac{t_s^{\alpha}}{\sigma(t_s)}\right)^2
+\left(\frac{n_b^{\alpha}}{\sigma(n_b)}\right)^2
+\left(\frac{t_b^{\alpha}}{\sigma(t_b)}\right)^2\right], \\
\Delta \chi_{para}^2&=&\sum_i \left(\frac{X^i-X^i_{true}}{\sigma(X^i)}\right)^2,
\end{eqnarray}
by using $N_{QE i}$ and $N_{CC}$, where $\alpha$ in 
$\Delta \chi^2_{sys}$ represents the sum for QE and CC events and 
$i$ in $\Delta \chi^2_{para}$ represents the sum for the parameters.
Namely, $X_i$ corresponds to the mixing angles, the mass squared differences and 
the matter density.
As the true value of the CP phase, we take $\delta_{true}=90^{\circ}$ and $180^{\circ}$ 
and plot the value of $\Delta \chi^2$ as the function of test value in figure 4.
We use the Globes software \cite{globes} in this calculation.

\FIGURE[!t]{
\begin{tabular}{cc}
\resizebox{72mm}{!}{\includegraphics{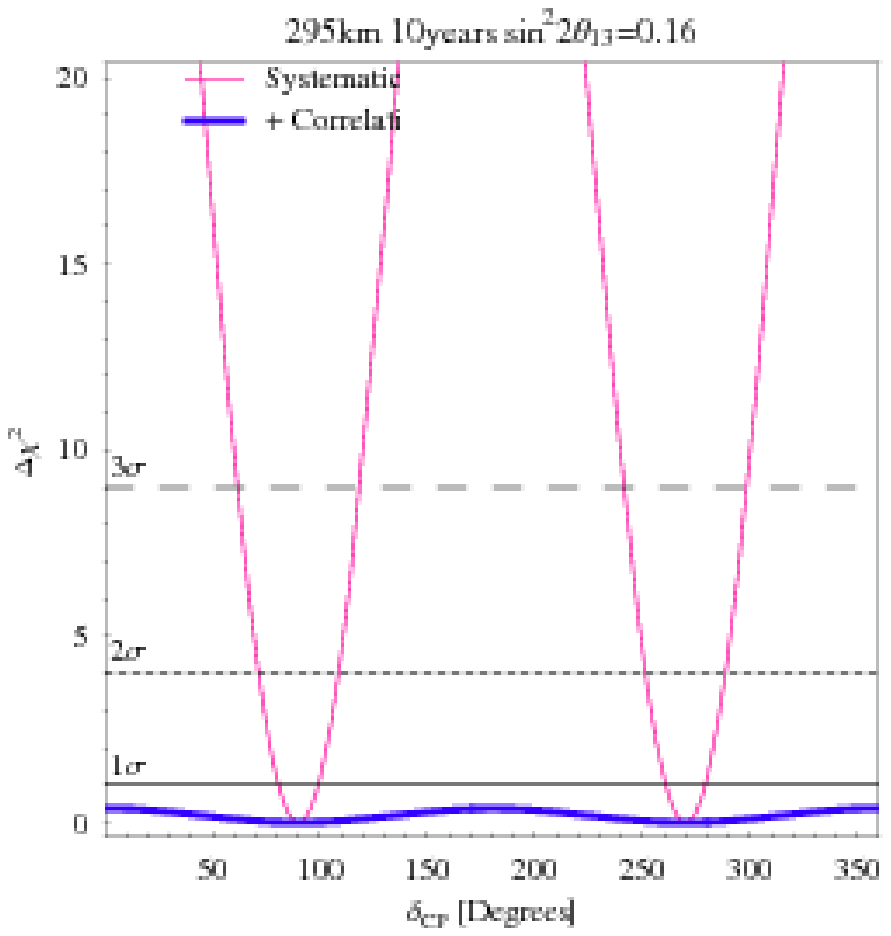}} &
\hspace{-0.5cm}
\resizebox{72mm}{!}{\includegraphics{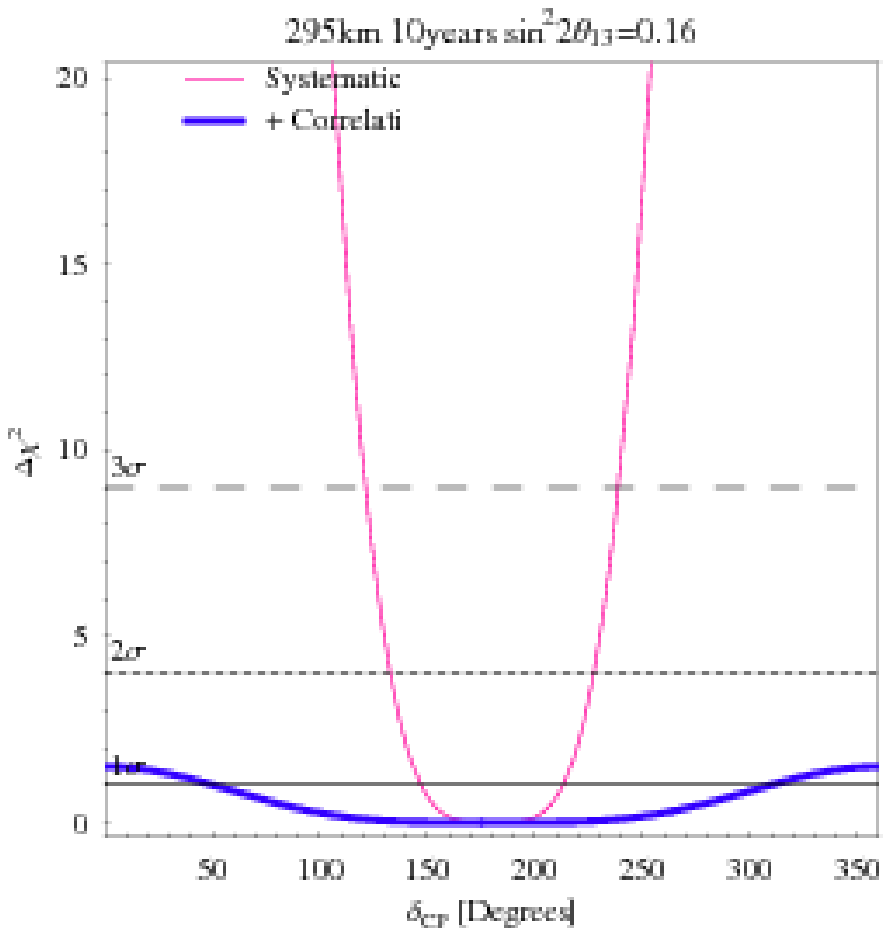}} \\
\resizebox{72mm}{!}{\includegraphics{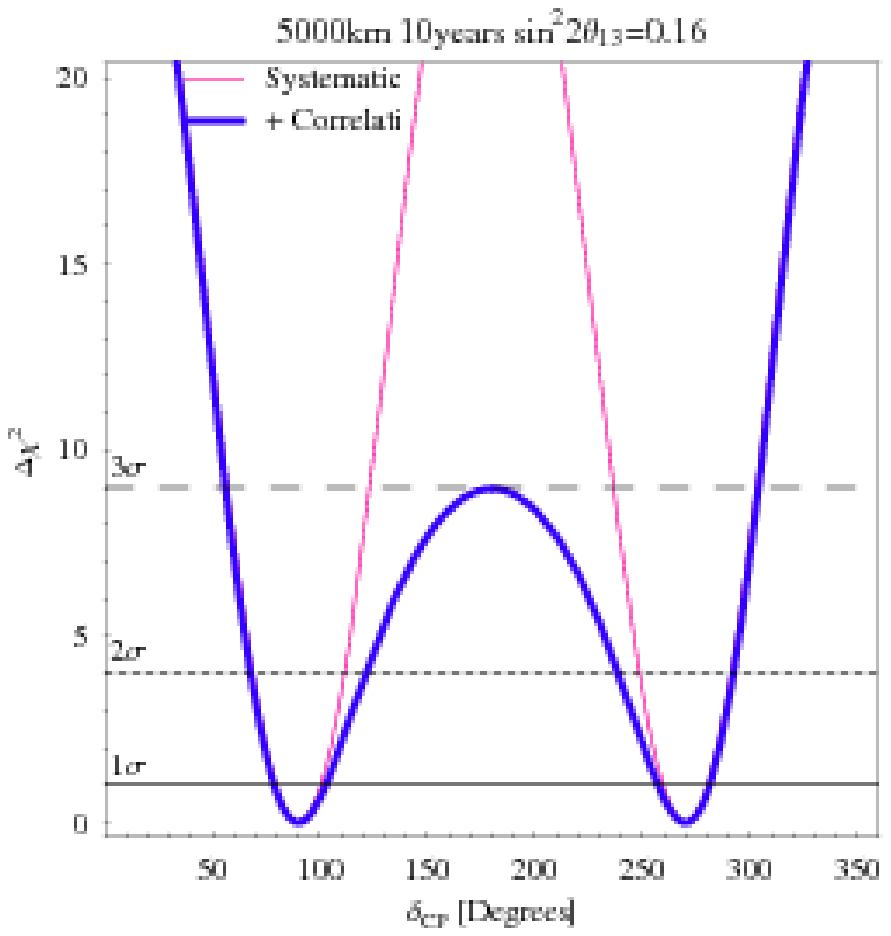}} &
\hspace{-0.5cm}
\resizebox{72mm}{!}{\includegraphics{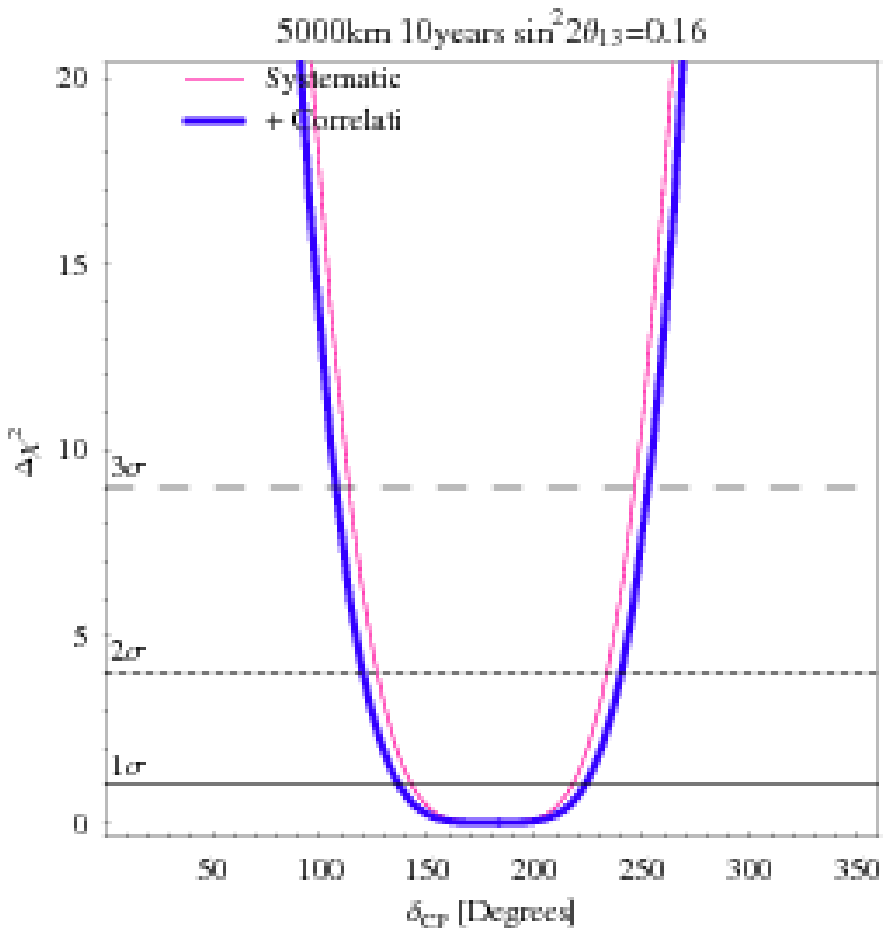}}
\end{tabular}
  \caption{\label{fig:fig4}%
$\Delta \chi^2$ by assuming the
true values, $\delta_{true}=90^{\circ}, 180^{\circ}$.
The pink line represents $\Delta \chi^2$ calculations including only systematics 
and the blue including also parameter uncertainties. 
Left and right figures correspond 
to $\delta_{true}=90^{\circ}$ and $180^{\circ}$. 
Top and bottom figures are calculated under the assumption of normal hierarchy 
in the baseline length $L=295$km and $5000$km.
}}
In figure 4, left and right figures correspond to the case of 
$\delta_{true}=90^{\circ}$ and $180^{\circ}$.
Top and bottom figures show the $\Delta \chi^2$ calculated in the baseline 
$L=295$km and $5000$km for the case of normal hierarchy.
{}From the top figures, we can find 
that the measurement of the CP phase 
is difficult in $L=295$km because of the uncertainties of parameters.
On the other hand, the difficulties are largely decreased 
for the case of $L=5000$km.
We can find from figure 4 that the allowed range is $50^{\circ}$ ($106^{\circ}$) 
in 1-$\sigma$ (2-$\sigma$) C.L. for $\delta_{true}=90^{\circ}$ and 
$88^{\circ}$ ($120^{\circ}$) in 1-$\sigma$ (2-$\sigma$) C.L. for 
$\delta_{true}=180^{\circ}$. 
We do not show the figures for the case of inverted hierarchy, 
because they are almost similar to the case of 
normal hierarchy
and only the CP sensitivity becomes slightly worse.
This can be understood by using the approximate formula of 
$A_{\mu\mu}$ given in (\ref{Ammapp}) as follows.
Namely, in low energy region, $A_{\mu\mu}$ does not depend on 
the sign of $\Delta m_{31}^2$ because of the averaging of $A_2$ 
and in high energy region, 
the number of events is suppressed due to the MSW effect compared to 
the case of normal hierarchy through $A_2$.

Next, we investigate how the CP sensitivity depends on 
the baseline length and the magnitude of $\theta_{13}$.
The value of CP sensitivity is shown in figure 5 and figure 6.
The CP sensitivity stands for how 
wide CP angles within $360^{\circ}$ 
is allowed at certain C.L. when we set a true value of the CP phase.
In the case that the allowed range is narrow, we can determine the CP phase precisely.
Left and right figures correspond to the case of $\delta_{true}=90^{\circ}$ and $180^{\circ}$.
Here, only the case for normal hierarchy is shown as we obtain similar result 
for inverted hierarchy.

\FIGURE[!t]{
\begin{tabular}{cc}
\hspace{-0.8cm}
\resizebox{82mm}{!}{\includegraphics{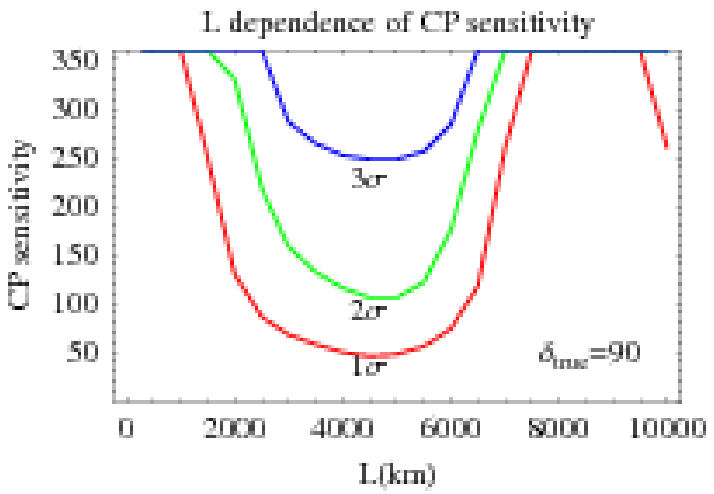}} &
\hspace{-0.8cm}
\resizebox{82mm}{!}{\includegraphics{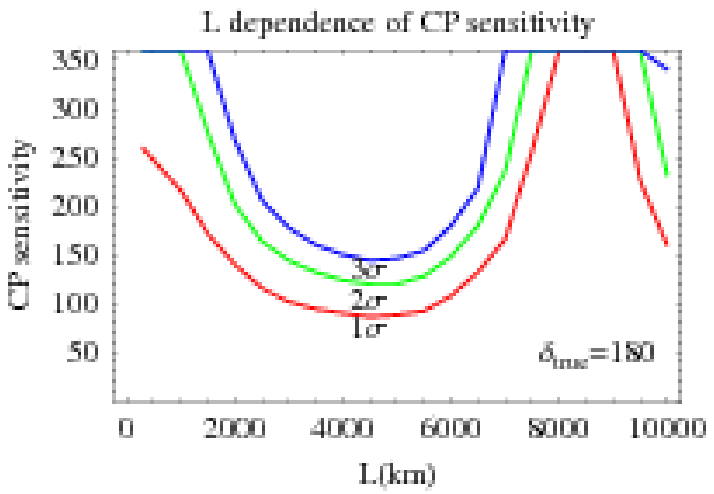}} 
\end{tabular}
  \caption{\label{fig:fig5}%
$L$ dependence of CP sensitivity. 
In left and right figures, we take $\delta_{true}=90^{\circ}$ and $180^{\circ}$.
Red, green and blue lines show 1,2 and 3-$\sigma$ C.L. lines 
respectively. We use the same parameters as in figure 4.
}}

Figure 5 shows the baseline dependence of the CP sensitivity.
In the both cases, the best sensitivity is realized around $L=4500\sim 5000$km.
This coincides well with the result obtained in (\ref{25}).
In the baseline length of $L=10000$km, 
the CP sensitivity is not good because of the small statistics due to the 
too long distance. 

Figure 6 shows the $\theta_{13}$ dependence of the CP sensitivity at $L=5000$km.
We can see that the CP sensitivity becomes worse gradually according to the 
decrease of $\theta_{13}$ in both cases.
This can be understood by eq.(\ref{Ammapp}) as $A_{\mu\mu}$ is proportional 
to $s_{13}$.
It is also found that the CP sensitivity for $\delta_{true}=180^{\circ}$ 
is good compared with $\delta_{true}=90^{\circ}$ in 3-$\sigma$ C.L.
This is interpreted as follows.
In $\nu_{\mu}\to \nu_{\mu}$ oscillations, the probability depends on the CP phase 
through $\cos \delta$.
So, we obtain the largest differences between the probabilities 
for the two extremes, $\delta=0^{\circ}$ and $\delta=180^{\circ}$, 
and the probability for $\delta=90^{\circ}$ 
is in between these two cases.
This reduces the difference to other probabilities 
and the allowed range becomes wide.

Next, let us consider the dependence of CP sensitivity on the systematic 
errors.
As the backgrounds in $\nu_{\mu}$ disapearance channel are small compared 
to the signal, $\sigma(n_b)$ and $\sigma(t_b)$ hardly change the results.
The main factor affecting the results is the signal normalization $\sigma(n_s^{CC})$.

\FIGURE[!t]{
\begin{tabular}{cc}
\hspace{-1.1cm}
\resizebox{86mm}{!}{\includegraphics{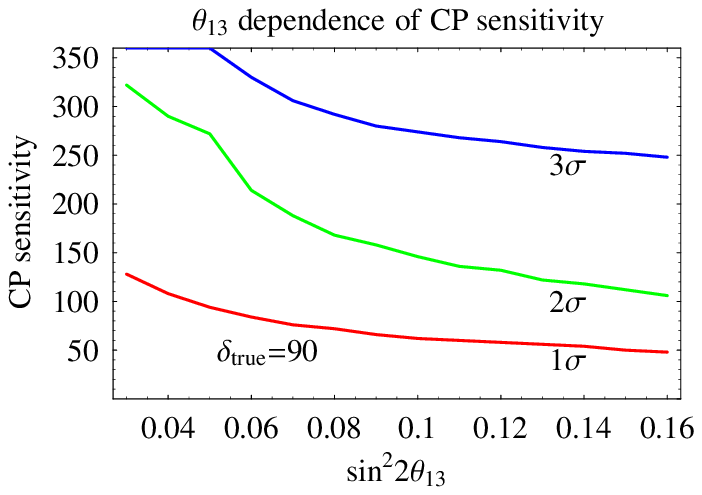}} &
\hspace{-1.1cm}
\resizebox{86mm}{!}{\includegraphics{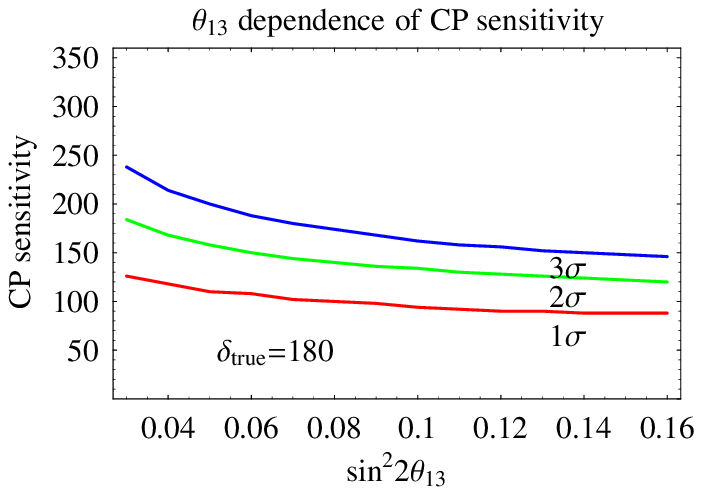}} 
\end{tabular}
  \caption{\label{fig:fig6}%
$\theta_{13}$ dependence of CP sensitivity at $L=5000$km. 
In left and right figures, we take $\delta_{true}=90^{\circ}$ and $180^{\circ}$.
Red, green and blue lines show 1,2 and 3-$\sigma$ C.L. lines 
respectively. 
}}

\FIGURE[!t]{
\begin{tabular}{cc}
\hspace{-1.1cm}
\resizebox{86mm}{!}{\includegraphics{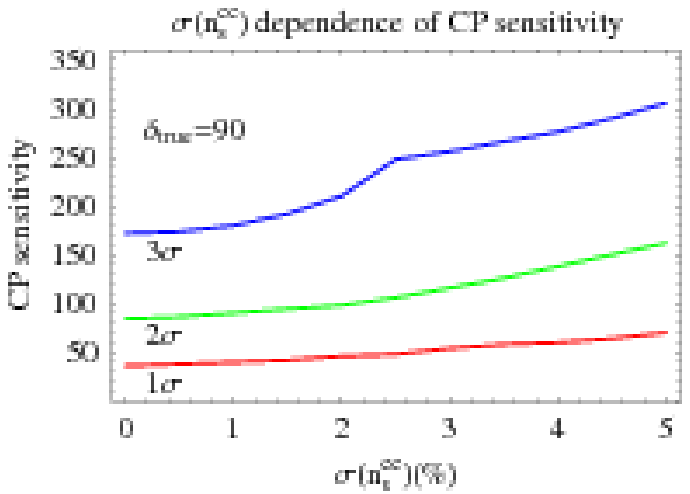}} &
\hspace{-1.1cm}
\resizebox{86mm}{!}{\includegraphics{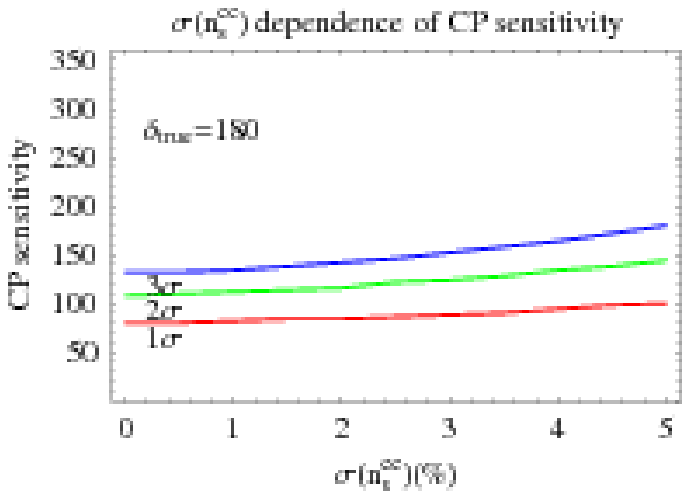}} 
\end{tabular}
  \caption{\label{fig:fig7}%
$\sigma(n_s^{CC})$ dependence of CP sensitivity at $L=5000$km. 
In left and right figures, we take $\delta_{true}=90^{\circ}$ and $180^{\circ}$.
Red, green and blue lines show 1,2 and 3-$\sigma$ C.L. lines 
respectively.
}}

Figure 7 shows the dependence of the CP sensitivity on $\nu_{\mu}$ beam normalization 
at $L=5000$km.
We can see that the CP sensitivity becomes worse gradually according to the 
increase of $\sigma(n_s^{CC})$ in both cases.
In this baseline, the value of CP phase is mainly determined by the total events 
averaged over the energy. 
As the total events are proportional to the $\cos \delta$ through $J_r$ in eq.(\ref{Ammapp}), 
it is considered to be difficult to distinguish the effect of $\delta$ 
and the signal normalization.

In the end of this section, we consider the reason why the CP sensitivity 
is so bad for the case of $L=295$km.
For the case of the short baseline, we could in principle determine 
the value of $\delta$ with high accuracy if the parameters, except $\delta$, 
are precisely determined because of the large statistics.
As the flux is inversely proportional to $L^2$, the statistics for $L\,=\,295$km 
becomes about 250 times larger than those for $L\,=\,5000$km.
However, the advantage is lost if the other 
parameters are including a few percent uncertainties.
This is due to the small contribution of the CP phase effect 
to the probability and the CP phase effect is hidden by the large contribution 
from other parameters.
The next question 
is which parameter prevents the 
determination of $\delta$.
It is found that the uncertainty of $\Delta m_{31}^2$ is related to the 
precise determination of $\delta$ by the numerical calculation as suggested 
in \cite{Donini0512}.
In figure 8, we show the dependence of the CP sensitivity at $L=295$km as the function of 
the uncertainty of $\Delta m_{31}^2$ in the case of normal hierarchy.
The experimental setup is assumed as the same as that in figure 4.
The left and right figures correspond to the cases for 
$\delta_{true}=90^{\circ}$ and $180^{\circ}$. 
Horizontal axis represents the $\delta(\Delta m_{31}^2)/\Delta m_{31}^2$ in 
the unit of \%, and the vertical axis represents the value of CP sensitivity.
{}From these figures, it is found that the CP sensitivity becomes gradually good 
below an uncertainty of $0.6\%$.

\FIGURE[!t]{
\begin{tabular}{cc}
\hspace{-1.1cm}
\resizebox{87mm}{!}{\includegraphics{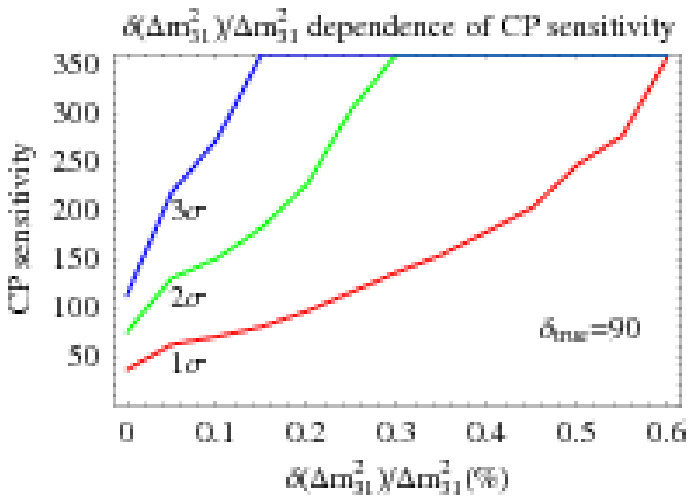}} &
\hspace{-1.3cm}
\resizebox{87mm}{!}{\includegraphics{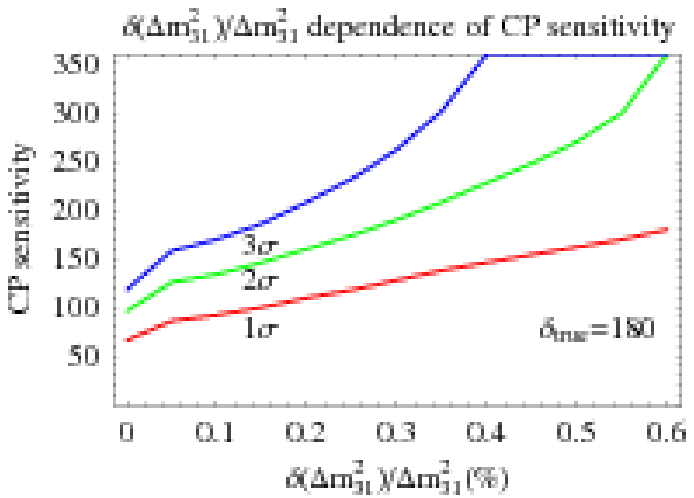}} 
\end{tabular}
  \caption{\label{fig:fig8}%
$\delta(\Delta m_{31}^2)$ dependence of CP sensitivity at $L=295$km. 
In left and right figures, we take $\delta_{true}=90^{\circ}$ and $180^{\circ}$.
Red, green and blue lines show 1,2 and 3-$\sigma$ C.L. lines 
respectively.
}}

\section{Correlation between $\delta$ and the Uncertainty of $\Delta m_{31}^2$}
In the previous section, it was shown that the uncertainty of $\Delta m_{31}^2$ 
becomes the serious obstacle in determining $\delta$ precisely.
In this section, we investigate these correlation in more detail 
by using the analytical expression and explore the possibility for the improvement.

For the case of $O(\Delta_{21}^{\prime})\ll 1$, the probability $P_{\mu\mu}$ is 
reduced to the well known expression 
\begin{eqnarray}
P_{\mu\mu}&\simeq& 1-\sin^2 \Delta_{31}^{\prime}-
\frac{8J_r\Delta_{21}}{\tilde{\Delta}_{21}}\sin \tilde{\Delta}_{21}^{\prime}
\sin \Delta_{31}^{\prime} \cos \Delta_{31}^{\prime}\cos \delta, \label{Ammapp2}
\end{eqnarray}
as shown in appendix.
If we perform the replacements 
$\Delta_{31} \to \Delta_{31}(1+\epsilon)$ and $\delta \to \delta^{\prime}$, 
the probability changes as 
\begin{eqnarray}
P_{\mu\mu}^{\prime}&\simeq&1-\sin^2 \Delta_{31}^{\prime}
-\left(\frac{8J_r\Delta_{21}}{\tilde{\Delta}_{21}}\sin \tilde{\Delta}_{21}^{\prime}
\cos \delta^{\prime}+\Delta_{31}\epsilon L\right)
\sin \Delta_{31}^{\prime} \cos \Delta_{31}^{\prime} \\
&=&1-\sin^2 \Delta_{31}^{\prime}
-\frac{8J_r\Delta_{21}}{\tilde{\Delta}_{21}}\sin \tilde{\Delta}_{21}^{\prime}
\sin \Delta_{31}^{\prime} \cos \Delta_{31}^{\prime}
\left(\cos \delta^{\prime}+\frac{\Delta_{31}\epsilon \tilde{\Delta}_{21}L}
{8J_r\Delta_{21}\sin \tilde{\Delta}_{21}^{\prime}}\right)
\end{eqnarray}
up to the leading order of $s_{13}$ and $\epsilon$. 
Here, we should note that the energy dependence of the correction term from 
the uncertainty of $\Delta m_{31}^2$ is approximately equal to that of the 
$\cos \delta^{\prime}$ term.
Namely, we cannot distinguish two probabilities 
$P_{\mu\mu}(\Delta_{31}(1+\epsilon), \delta^{\prime})$ and 
$P_{\mu\mu}(\Delta_{31}, \delta)$ when the relation 
\begin{eqnarray}
\cos \delta^{\prime}\simeq \cos \delta-\frac{\Delta_{31}\epsilon \tilde{\Delta}_{21} L}
{8J_r\Delta_{21}\sin \tilde{\Delta}_{21}^{\prime}}
\simeq \cos \delta-\frac{\Delta m_{31}^2\epsilon aL}
{8J_r\Delta m_{21}^2\sin \frac{aL}{2}} \label{38}
\end{eqnarray}
is satisfied, where we use the approximation $\tilde{\Delta}_{21}\simeq a$.
This means that the value of the CP phase cannot be determined if 
$\epsilon$ is larger than a certain value.
This is the correlation between $\delta$ and the uncertainty of $\Delta m_{31}^2$ 
and is a serious obstacle for measuring the CP phase in $\nu_{\mu} \to \nu_{\mu}$ oscillations.
The small value of the CP sensitivity in the baseline $L=295$km is due to this correlation.

Next, let us estimate the magnitude of $\epsilon$ giving the small CP sensitivity.
If we substitute $a =7.56\cdot 10^{-5}\rho Y_e$, $\rho=3.3$g/cm$^3$, $Y_e=0.494$, 
$J_r=s_{12}c_{12}s_{23}c_{23}s_{13}c_{13}^2\simeq 0.045$, 
$\Delta m_{31}^2=2.5\cdot 10^{-3}$eV$^2$ and $\Delta m_{21}^2=8.1\cdot 10^{-5}$eV$^2$ 
into (\ref{38}), we obtain 
\begin{eqnarray}
\cos \delta^{\prime}\simeq \cos \delta-\frac{5.3\cdot 10^{-2}\epsilon L}
{\sin \left(3.1\cdot 10^{-4}L\right)}. \label{relation}
\end{eqnarray}
In the case of relatively short baseline, the above relation is further reduced to 
\begin{eqnarray}
\cos \delta^{\prime}\simeq \cos \delta-170\epsilon \label{40}
\end{eqnarray}
and does not depend on the baseline length.
Let us illustrate the meaning of this relation in figure 9.

\FIGURE[!t]{
\begin{tabular}{cc}
\hspace{-0.5cm}
\resizebox{74mm}{!}{\includegraphics[scale=0.36,angle=-90]{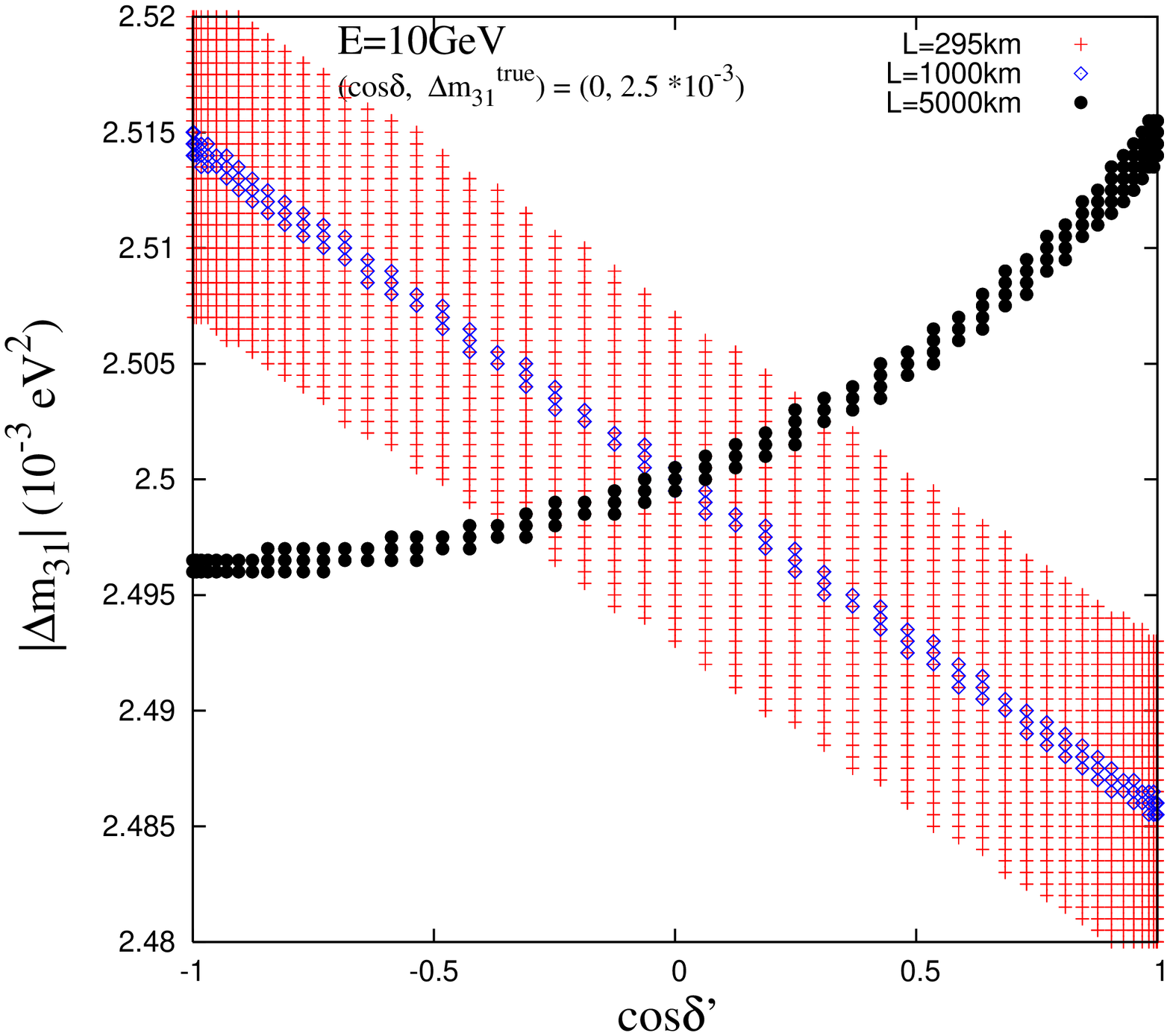}} &
\hspace{0.2cm}
\resizebox{74mm}{!}{\includegraphics[scale=0.36,angle=-90]{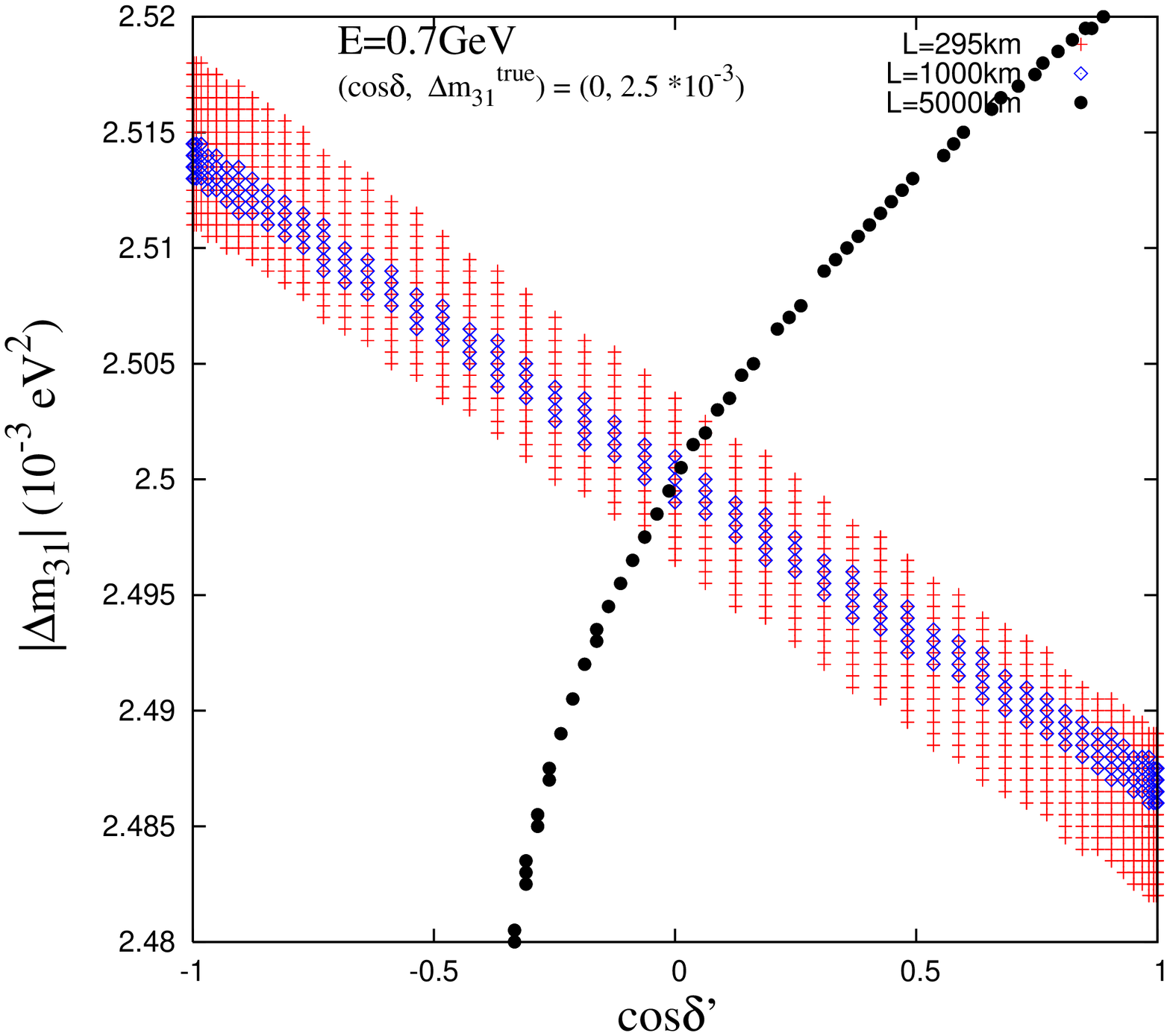}} 
\end{tabular}
  \caption{\label{fig:fig9}%
Correlation of $\delta$ with the uncertainty of $\Delta m_{31}^2$. 
In left and right figures, we take $E=10$ GeV and $0.7$ GeV.
Regions with same color have almost the same probability given by 
$|P_{\mu\mu}^{true}-P_{\mu\mu}|<0.001$ $(0.00005)$ in the left (right) figure.
Different colors correspond to the different baseline lengths.
}}

In figure 9, regions with same color have almost the same probability as that 
for $\delta_{true}=90^{\circ}$. 
More concretely, the region with $|P_{\mu\mu}-P_{\mu\mu}^{true}|\leq 0.001$ $(0.00005)$ 
is plotted in the left (right) figure.
The horizontal axis and vertical axis are taken as $\cos \delta^{\prime}$ and 
$\Delta m_{31}^2$ ($\epsilon$) respectively.
Left and right figures represent the case for $E=10$GeV and $0.7$GeV.
Red, blue and black colors correspond to the cases for 
$L=295$km, $1000$km and $5000$km.
It is obvious that the value of $\delta$ can be determined by measuring 
the probabilities for two different energies for the case of $L=5000$km 
from figure 9, because a superposition of the black curves for the 
two different energies (left and right figures) would lead to a clear 
intersection point representing the allowed region.
In contrary, we cannot determine the value of $\delta$ for a relatively 
short baseline like $L=295$km and $1000$km even if the probabilities 
are measured for two different energies, because of an almost identical 
overlap of regions with the same color for the two different energies 
(left and right figures).
The overlapping is over the whole range of angles $0^{\circ}\leq \delta 
\leq 360^{\circ}$.
We can determine the slope of these two regions as about $-0.006$, 
which is almost equal to the coefficient of $\epsilon$, namely $-1/170$ in (\ref{40}).
{}From this observation, we conclude that the uncertainty of $\Delta m_{31}^2$ 
of more than $0.6\%$ prevents us from determining $\delta$ from 
$\nu_{\mu}\to \nu_{\mu}$ oscillations only in relatively short baselines.
Or in other words, for the case of short baseline length the determination of 
the CP phase $\delta$ will become possible, 
if the uncertainty can be decreased below $0.6\%$, 
as demonstrated in the following section.

\section{CP Sensitivity with More Small Uncertainty}
In this section, we discuss the possibility for diminishing of parameter 
uncertainties including $\Delta m_{31}^2$ in order to measure $\delta$. 
We also investigate how the CP sensitivity is improved in a relatively short 
baseline superbeam experiment after diminishing of parameter uncertainties.
It is discussed that the uncertainty of $\Delta m_{21}^2$ can be 
reduced up to $3\%$ at 3-$\sigma$ by loading $0.1\%$ Gd in the detector 
of SK \cite{Choubey} and that the uncertainty 
of $\sin^2 \theta_{12}$ can be reduced up to $2\%$ at 1-$\sigma$ 
by using the reactor experiment with the baseline of $L\sim 60$km 
\cite{Bandyopadhyay}.
{}From these analysis, there is a possibility for diminishing the uncertainties 
up to $1\%$ at 1-$\sigma$ C.L. for both $\Delta m_{21}^2$ and $\theta_{12}$ 
in near future experiments.
The uncertainty of $\Delta m_{31}^2$ can be reduced up to $1\%$ at the T2K experiment 
and the NO$\nu$A experiment.
However, it is required that the uncertainty has to be reduced one more order 
of magnitude in order to receive the sensitivity for the CP phase 
as discussed in the previous section.
So, we consider two baselines.
We measure the value of $\Delta m_{31}^2$ as precisely as possible in one baseline  
and then measure the value of $\delta$ by using only $\nu_{\mu}\to \nu_{\mu}$ oscillations 
in another baseline.
As one of the real models, we consider the T2KK experiment \cite{T2KK}.
In this experiment, the neutrinos emitted from Tokai are observed in two detectors at 
Kamioka and Korea. 
One of the aims in the T2KK experiment is to decrease the systematic error by using the 
same beam, but we use this experiment to reduce the uncertainty of $\Delta m_{31}^2$.
Here, we assume that we perform the precise measurement of $\Delta m_{31}^2$ 
in Korea WC detector at $L=1000$km with $500$kt fiducial mass and 
then we measure the value of $\delta$ in Kamioka WC detector at $L=295$km with the same 
size.
The size of detectors assumed here is larger than those considered in \cite{T2KK}.

At first, we set the true value of $\Delta m_{31}^2$ as $2.5\cdot 10^{-3}$eV$^2$ 
and calculate $\Delta \chi^2$ as the function of test value at $L=1000$km.
Here, we consider the case of normal hierarchy. 
In the case of inverted hierarchy, we obtain the similar features.
In our analysis, we consider $\delta$ as free parameter and set the uncertainties of 
$1\%$ for $\Delta m_{21}^2$ and $\theta_{12}$.
We use the same uncertainties as in sec.3 for other parameters.
Figure 10 shows the value of $\Delta \chi^2$.
Pink and blue lines correspond to the case of 
$\delta_{true}=90^{\circ}$ and $180^{\circ}$.
We take into account not only the systematic error but also the uncertainties of 
various parameters in calculating $\Delta \chi^2$.
{}From this figure, the test values separated from the true value more than $0.2$-$0.3\%$ 
can be excluded at 1-$\sigma$ C.L. with 2 d.o.f.

Figure 11 shows the value of $\Delta \chi^2$ as the function of test value $\delta$, 
fixing the uncertainty of $\Delta m_{31}^2$ to $0.2\%$ at 1-$\sigma$.
About uncertainties of other parameters, we use the same as in figure 10.

We found that there is the CP sensitivity even at relatively short baseline like 
$L=295$km due to the decrease of parameter uncertainties.
For $\delta_{true}=90^{\circ}$, the allowed range is $86^{\circ}$ ($190^{\circ}$) 
at 1-$\sigma$ (2-$\sigma$) and for $\delta_{true}=180^{\circ}$ the allowed 
range is $100^{\circ}$ ($148^{\circ}$) at 1-$\sigma$ (2-$\sigma$).
In such short baseline, 
we have an advantage in the statistical point of view.
Therefore, this strategy may be better if we can determine the parameters 
precisely in future experiments.

In figure 12, we show the $\theta_{13}$ dependence of the CP sensitivity 
in the baseline $L=295$km 
when $\delta(\Delta m_{31}^2)/\Delta m_{31}^2$ is fixed at $0.2\%$.
The experimental setup is the same as in sec.3.

\FIGURE[!t]{
\begin{tabular}{c}
\resizebox{100mm}{!}{\includegraphics{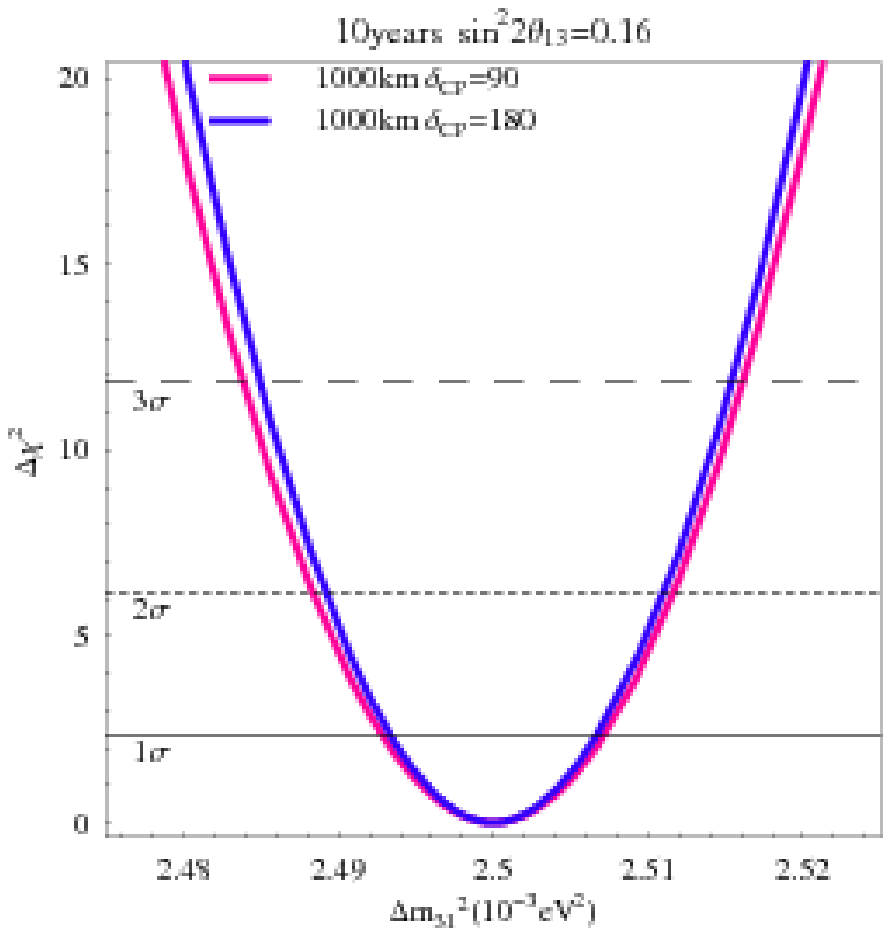}}
\end{tabular}
  \caption{\label{fig:fig10}%
Precision of $\Delta m_{31}^2$. Blue and pink lines correspond to 
the case of $\delta_{true}=90^{\circ}$ and $180^{\circ}$.
$\Delta \chi^2$ is calculated considering both systematics and parameter uncertainties.
}}

{}From figure 12, we can see that the CP sensitivity becomes worse mildly 
according to the decrease of $\theta_{13}$ as in figure 6.
In the case of $\delta_{true}=90^{\circ}$ and $180^{\circ}$, the allowed range is about 
$222^{\circ}$ and $160^{\circ}$ respectively at 1-$\sigma$ 
for the value of $\sin^2 2\theta_{13}=0.03$.

Finally, let us comment the dependence of CP sensitivity on the systematic errors.
In the baseline of $L=295$km, the first term $A_1$ in eq.(\ref{Ammapp}) can be neglected 
and only the second term $A_2$ contributes to the determination of $\delta$.
Namely, we obtain the information of $\delta$ by the energy dependence of 
QE $\nu_{\mu}$ events.
In our analysis, we take the signal normalization of QE $\nu_{\mu}$ 
events as free from the beginning.
Therefore, the signal normalization hardly affect the CP sensitivity in this baseline.
This is in contrast to the case of $L=5000$km.

\FIGURE[!t]{
\begin{tabular}{cc}
\resizebox{75mm}{!}{\includegraphics{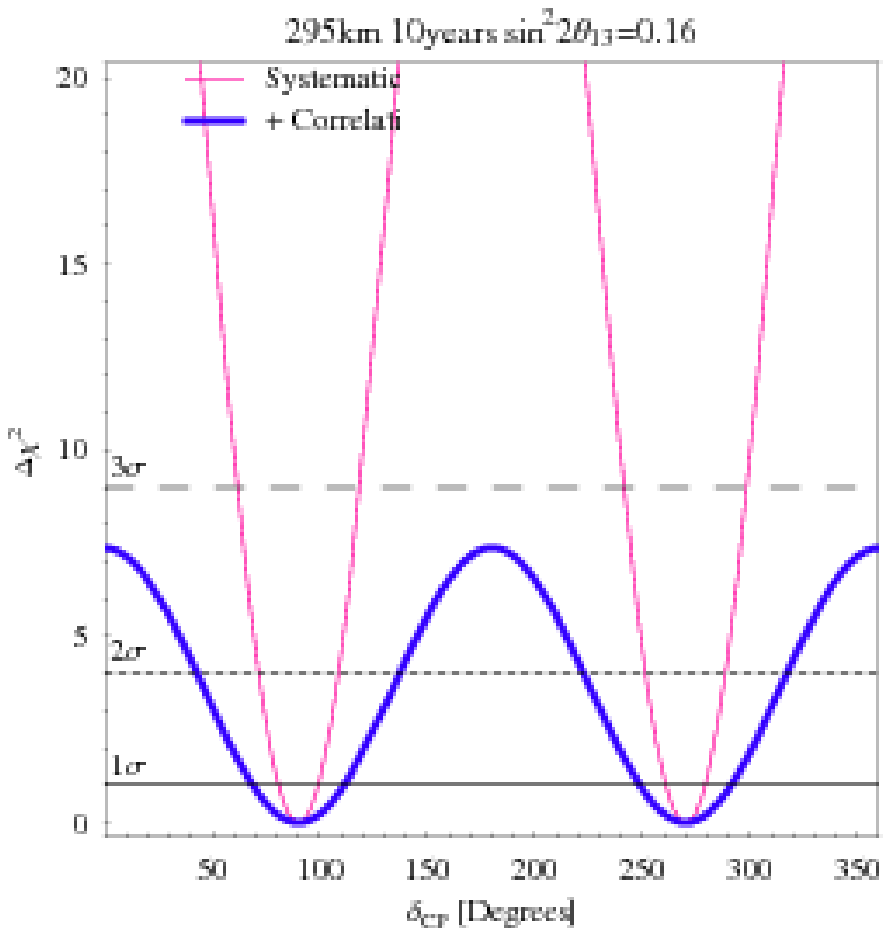}} &
\hspace{-1cm}
\resizebox{75mm}{!}{\includegraphics{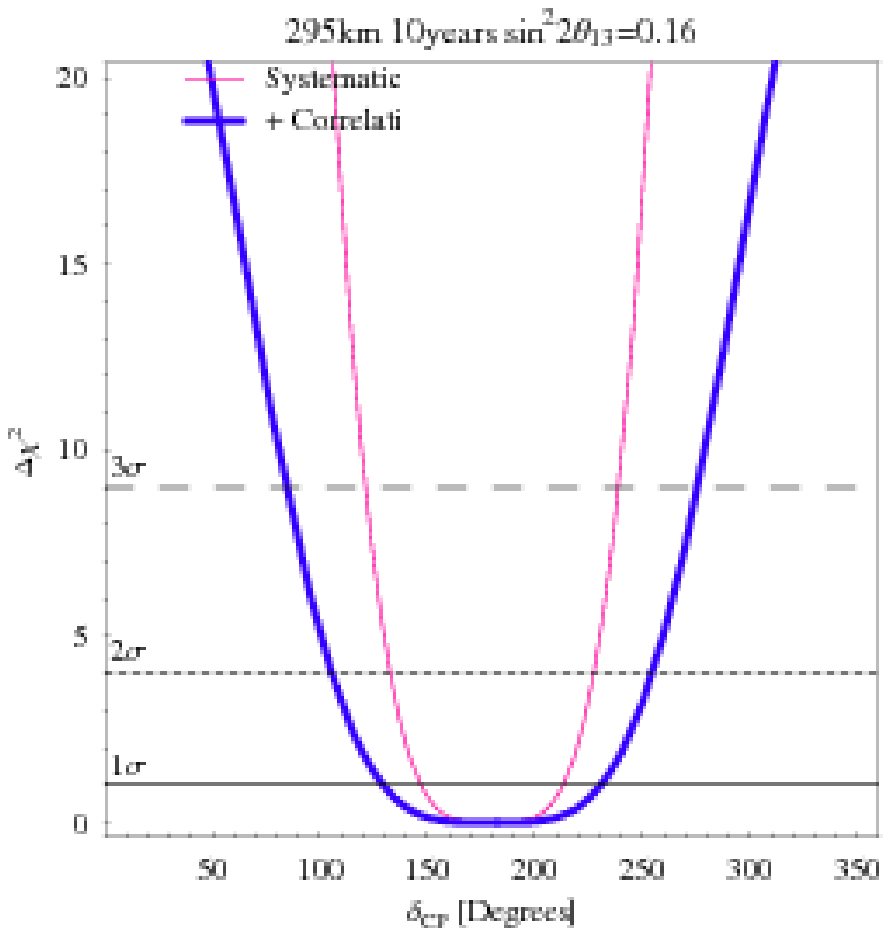}} \\
\end{tabular}
  \caption{\label{fig:fig11}%
$\Delta \chi^2$ by assuming the
true values, $\delta=90^{\circ}, 180^{\circ}$.
Pink and blue lines represent the $\Delta \chi^2$ considering only systematics 
and including also parameter uncertainties.
}}

\FIGURE[!t]{
\begin{tabular}{cc}
\hspace{-1.1cm}
\resizebox{86mm}{!}{\includegraphics{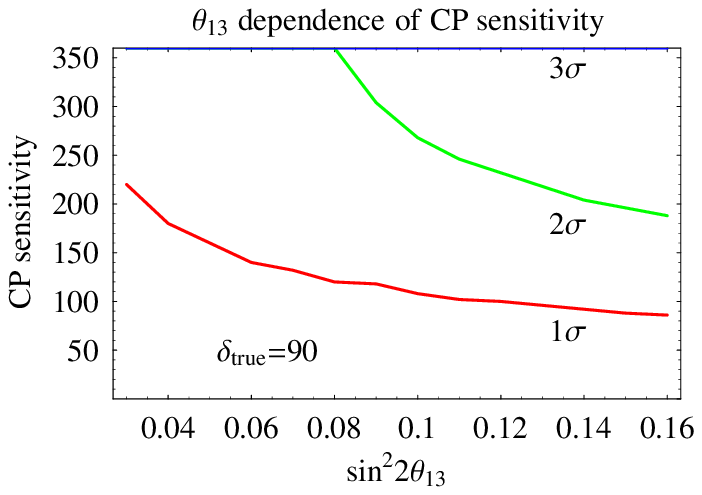}} &
\hspace{-1.1cm}
\resizebox{86mm}{!}{\includegraphics{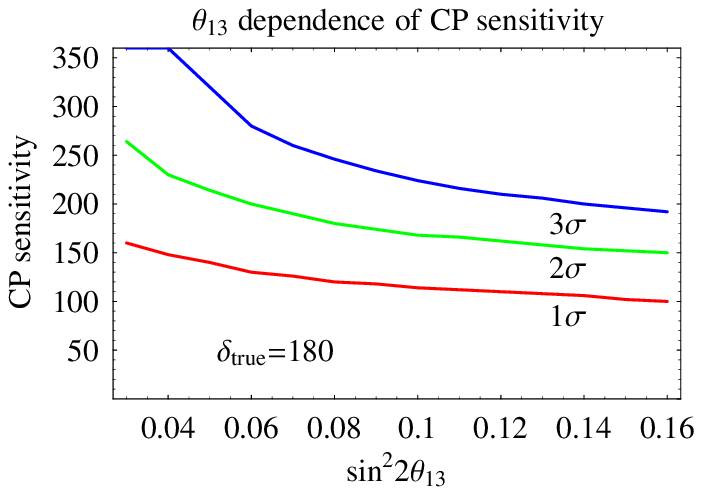}} 
\end{tabular}
  \caption{\label{fig:fig12}%
$\theta_{13}$ dependence of CP sensitivity. 
In left and right figures, we take $\delta_{true}=90^{\circ}$ and $180^{\circ}$.
Red, green and blue lines show 1,2 and 3-$\sigma$ C.L. lines 
respectively.
}}

\section{Summary and Discussion}
\label{sec:summary}
In this paper, we have explored new possibilities for experiments to be 
performed in the next decade for the case that 
$\sin^2 2\theta_{13}$ is larger than 
$0.03$. 
The $\theta_{13}$ will be determined in the next generation reactor 
experiments.
Concretely, we have investigated whether the CP phase can be measured by only 
$\nu_{\mu}\to \nu_{\mu}$ oscillations independently of 
$\nu_{\mu} \to \nu_e$ oscillations.
If we can measure the CP phase in two different channels independently 
and there is a difference for these values, this would be considered
as an evidence of new physics.
Below, the results obtained in this paper are listed.

\begin{itemize}
\item At first, we have investigated the energy and the baseline 
where the $\cos \delta$ term included in $P_{\mu\mu}$ 
becomes large by using both numerical and analytical methods.
As the result, we found from (\ref{2.21}) and (\ref{25}) 
that the coefficient $A_{\mu\mu}$ has its largest value 
around $E=0.43$GeV and $L=5000$ and $10000$km 
in the earth mantle.
The difference of the probabilities attains the maximal value 
about $0.2$ due to the CP phase effect even after the 
averaging over the energy.

\item Next, we have considered the same beam and the detector 
as in the T2HK experiment but the baseline $L=295$km and $5000$km 
and have calculated the CP sensitivity by using $\chi^2$ method.
As the result, when the uncertainty of $\Delta m_{31}^2$ is larger than $1\%$, 
the CP sensitivity has its best value for the baseline around $L=5000$km.
The allowed range becomes $50^{\circ}$ ($88^{\circ}$) at 1-$\sigma$ in $L=5000$km 
when $\delta_{true}=90^{\circ}$ ($180^{\circ}$).
On the other hand, we have almost no sensitivity for the CP phase 
in the case of $L=295$km because of the effect of parameter uncertainties.

\item We have shown both numerically and analytically 
that the uncertainty of $\Delta m_{31}^2$ 
is particularly important in determining the value of $\delta$.
If the uncertainty of $\Delta m_{31}^2$ will be less than $0.6\%$, 
a certain sensitivity for $\delta$ will be obtained even in the 
relatively short baseline like $L\leq 1000$km.

\item We have explored the possibility for measuring the CP phase in 
relatively short baseline by diminishing the uncertainty of 
$\Delta m_{31}^2$.
As concrete experimental setup, we have considered the T2KK experiment.
We used one baseline from Tokai to Korea in order to determine 
$\Delta m_{31}^2$ precisely and used the other baseline from Tokai 
to Kamioka to measure the CP phase by $\nu_{\mu}\to \nu_{\mu}$ oscillations.
As the result, the uncertainty of $\Delta m_{31}^2$ can be reduced up to 
about $0.2\%$ and the allowed range becomes about $222^{\circ}$ ($160^{\circ}$) 
at 1-$\sigma$ when $\delta_{true}=90^{\circ}$ ($180^{\circ}$) at 1-$\sigma$.
So, we have the possibility for measuring the CP phase although the sensitivity is not so good.
\end{itemize}

In future, the mixing angles and the mass squared differences are precisely 
measured in various kinds of experiments.
If $\theta_{13}$ is found in the next generation reactor experiments in addition to 
this improvement, it will be very important to consider the strategy 
for exploring the new physics.
As one of the strategies, the possibility for measuring the CP phase in 
$\nu_{\mu}\to \nu_{\mu}$ oscillation is interesting.
We will further investigate in our next work how the contribution of new physics 
appears in future experiments.

\acknowledgments

We would like to thank Prof. Wilfried Wunderlich
(Tokai university) for helpful comments and
advice on English expressions.
The work of T.Y. was supported by 21st Century COE Program of Nagoya 
University.

\appendix

\section{Derivation of Approximate Formula for $A_{\mu\mu}$}
\label{sec:appendix-a}
In this appendix, let us derive the approximate formula of $A_{\mu\mu}$ 
used in sec.2 for the case of constant matter.

At first, in the case of $\theta_{23}=45^{\circ}$ and the symmetric matter profile, 
the probability is given by 
\begin{eqnarray}
P_{\mu\mu}
=A_{\mu\mu}\cos \delta
+C_{\mu\mu}+D\cos 2\delta,
\end{eqnarray}
from the exact formula \cite{Yokomakura, Kimura0203}.
Here, we can neglect $D$ because of the higher order terms of small parameters.
Concrete expression for $A_{\mu\mu}$ \cite{Kimura0203} is given by 
\begin{eqnarray}
A_{\mu\mu}&=&\sum_{(ijk)}^{{\rm cyclic}}\frac{8[J_r\Delta_{21}
\Delta_{31}\lambda_k(\lambda_k-\Delta_{31})
-(A_{\mu\mu})_k]}
{\tilde{\Delta}_{jk}^2\tilde{\Delta}_{ki}^2}
\cos \tilde{\Delta}_{ij}^{\prime}
\sin \tilde{\Delta}_{jk}^{\prime}
\sin \tilde{\Delta}_{ki}^{\prime},
\label{23}
\end{eqnarray}
where 
\begin{eqnarray}
(A_{\mu\mu})_k&=&\Delta_{21}^2 J_r\times
\left[\Delta_{31}\lambda_k
(c_{12}^2-s_{12}^2)+\lambda_k^2 s_{12}^2
-\Delta_{31}^2 c_{12}^2 \right].  \label{7.51}
\end{eqnarray}

In the energy region around $E=1$GeV, 
we have $\lambda_1=O(\Delta_{21}), \lambda_2=O(\Delta_{21})$ and 
$\lambda_3\simeq \Delta_{31}$.
{}From these order counting, we can neglect $\lambda_1$ and $\lambda_2$ 
compared with $\lambda_3$.
On the other hand, we cannot neglect $\Delta_{21}^{\prime}$ included 
in the oscillating term for long baseline such that 
the relation $\Delta_{21}^{\prime}=\Delta m_{21}^2L/4E=O(1)$ is satisfied. 
If we leave only leading order terms in small parameters $\Delta_{21}$ and 
$s_{13}$, the expression for $A_{\mu\mu}$ is reduced to 
\begin{eqnarray}
A_{\mu\mu}
&\simeq& 
\frac{8J_r\Delta_{21}\Delta_{31}
[\lambda_1(\lambda_1-\Delta_{31})
+\Delta_{31}\Delta_{21}c_{12}^2]}
{\tilde{\Delta}_{31}^2\tilde{\Delta}_{21}^2}
\cos \tilde{\Delta}_{23}^{\prime}
\sin \tilde{\Delta}_{31}^{\prime}
\sin \tilde{\Delta}_{12}^{\prime} \nonumber \\
&+&\frac{8J_r\Delta_{21}\Delta_{31}
[\lambda_2(\lambda_2-\Delta_{31})
+\Delta_{31}\Delta_{21}c_{12}^2]}
{\tilde{\Delta}_{32}^2\tilde{\Delta}_{21}^2}
\cos \tilde{\Delta}_{31}^{\prime}
\sin \tilde{\Delta}_{12}^{\prime}
\sin \tilde{\Delta}_{23}^{\prime} \nonumber 
\\
&=& 
\frac{4J_r\Delta_{21}\Delta_{31}
[\lambda_1(\lambda_1-\Delta_{31})
+\Delta_{31}\Delta_{21}c_{12}^2]}
{\tilde{\Delta}_{31}^2\tilde{\Delta}_{21}^2}
\sin \tilde{\Delta}_{12}^{\prime}
\{\sin \tilde{\Delta}_{21}^{\prime}+
\sin (\tilde{\Delta}_{31}^{\prime}-\tilde{\Delta}_{23}^{\prime})\}
\nonumber \\
&+&\frac{4J_r\Delta_{21}\Delta_{31}
[\lambda_2(\lambda_2-\Delta_{31})
+\Delta_{31}\Delta_{21}c_{12}^2]}
{\tilde{\Delta}_{32}^2\tilde{\Delta}_{21}^2}
\sin \tilde{\Delta}_{12}^{\prime}
\{\sin \tilde{\Delta}_{21}^{\prime}-
\sin (\tilde{\Delta}_{31}^{\prime}-\tilde{\Delta}_{23}^{\prime})\}
\nonumber \\
&=& 
-\frac{4J_r\Delta_{21}\Delta_{31}}{\tilde{\Delta}_{21}^2}
\sin \tilde{\Delta}_{21}^{\prime 2} \nonumber \\
&&\times\left[\frac{\lambda_1(\lambda_1-\Delta_{31})+\Delta_{31}\Delta_{21}c_{12}^2}
{\tilde{\Delta}_{31}^2}+
\frac{\lambda_2(\lambda_2-\Delta_{31})+\Delta_{31}\Delta_{21}c_{12}^2}
{\tilde{\Delta}_{32}^2}\right]\nonumber \\
&&+\frac{4J_r\Delta_{21}\Delta_{31}}{\tilde{\Delta}_{21}^2}
\sin \tilde{\Delta}_{12}^{\prime}
\sin (2\tilde{\Delta}_{31}^{\prime}-\tilde{\Delta}_{21}^{\prime}) \nonumber 
\end{eqnarray}
\begin{eqnarray}
&&\times\left[\frac{\lambda_1(\lambda_1-\Delta_{31})+\Delta_{31}\Delta_{21}c_{12}^2}
{\tilde{\Delta}_{31}^2}-
\frac{\lambda_2(\lambda_2-\Delta_{31})+\Delta_{31}\Delta_{21}c_{12}^2}
{\tilde{\Delta}_{32}^2}\right]
\nonumber \\
&\simeq& 
\frac{4J_r\Delta_{21}}{\tilde{\Delta}_{21}^2}
\sin \tilde{\Delta}_{21}^{\prime 2}
\left[\lambda_1-2\Delta_{21}c_{12}^2+\lambda_2\right]
-\frac{4J_r\Delta_{21}}{\tilde{\Delta}_{21}}\sin \tilde{\Delta}_{21}^{\prime}
\sin (2\tilde{\Delta}_{31}^{\prime}-\tilde{\Delta}_{21}^{\prime})
\nonumber \\
&\simeq& 
\frac{4J_r\Delta_{21}}{\tilde{\Delta}_{21}^2}
\sin \tilde{\Delta}_{21}^{\prime 2}
\left[a-\Delta_{21}\cos 2\theta_{12}\right]
-\frac{4J_r\Delta_{21}}{\tilde{\Delta}_{21}}\sin \tilde{\Delta}_{21}^{\prime}
\sin (2\Delta_{32}^{\prime}).
\end{eqnarray}
This is the derivation of (\ref{Ammapp}).
In this approximate formula, the MSW effect due to the 1-3 mixing angle 
is not included, so there is some differences from the exact one 
in high energy.
On the other hand, this approximate formula coincides well in low 
energy.
In this expression, if $L/E$ is small enough, the first term can be neglected 
and is reduced to 
\begin{eqnarray}
A_{\mu\mu}\simeq -\frac{4J_r\Delta_{21}}{\tilde{\Delta}_{21}}\sin \tilde{\Delta}_{21}^{\prime}
\sin (2\Delta_{32}^{\prime})
\simeq -\frac{8J_r\Delta_{21}}{\tilde{\Delta}_{21}}\sin \tilde{\Delta}_{21}^{\prime}
\sin (\Delta_{31}^{\prime})\cos (\Delta_{31}^{\prime}).
\end{eqnarray}
This is the third term of (\ref{Ammapp2}).


\providecommand{\href}[2]{#2}\begingroup\raggedright\endgroup

\end{document}